\documentclass[11pt,letterpaper]{article}
\usepackage[numbers,sort&compress]{natbib}

\usepackage{authblk}
\usepackage[english]{babel}

\usepackage{amsmath,amsfonts,amssymb,amsthm}
\usepackage{graphicx,graphbox}
\usepackage{float}

\usepackage{hyperref}
\hypersetup{colorlinks=true,linkcolor=black,
 anchorcolor = blue,
 citecolor = blue,
 filecolor = blue,
 urlcolor = blue}
\usepackage[left=1in,right=1in,top=1in,bottom=1in]{geometry}   
\usepackage{tocloft}

\usepackage{graphicx}
\usepackage{float}
\usepackage{amsmath,amssymb,amsfonts,mathrsfs}
\usepackage{xcolor}

\usepackage{caption}
\usepackage{subcaption}
\usepackage{comment}
\usepackage{enumitem} 

\usepackage[normalem]{ulem} 

\usepackage{dsfont}
\usepackage{bm}
\usepackage{bbold}
\usepackage{physics} 
\usepackage[percent]{overpic} 

\begin{document}

\title{
\textbf{Causal structure in spin-foams}\\[.5em]
}

\author{
Eugenio Bianchi,${}^{ab}\;$  and
Pierre Martin-Dussaud$\,{}^{c}\;$ 
}
\date{}

\maketitle

\begin{center}
\vspace{-2em}
{\footnotesize 
${}^{a}$ Department of Physics, The Pennsylvania State University, University Park, Pennsylvania 16802, USA\\	
${}^{b}$ Institute for Gravitation and the Cosmos, The Pennsylvania State University,  Pennsylvania 16802, USA\\
${}^{c}$ Basic Research Community for Physics e.V., Mariannenstra\ss e 89, Leipzig, Germany\\[.5em]
 }
\end{center}

\vspace{2em}

\abstract{The metric field of general relativity is almost fully determined by its causal structure. Yet, in spin-foam models for quantum gravity, the role played by the causal structure is still largely unexplored. The goal of this paper is to clarify how causality is encoded in such models. The quest unveils the physical meaning of the orientation of the two-complex and its role as a dynamical variable. We propose a causal version of the EPRL spin-foam model and discuss the role of the causal structure in the reconstruction of a semiclassical spacetime geometry.}

\section{Introduction}

The information carried by a space-time metric is mainly of a causal nature. Indeed, Malament's theorem states that the causal relations between the points of a 4d manifold fully determine the metric, up to a conformal factor given at each point \cite{Malament:1977aa,Hawking:1976fe,Hawking:1973a}. In quantum models of space-time, the role of the metric is usually played by more fundamental objects, like spins and intertwiners in spin-foam models \cite{Perez:2012wv,Bianchi:2017hjl,Livine:2024hhc} for the dynamics of loop quantum gravity \cite{Rovelli:2014a,Thiemann:2007pyv,Ashtekar:2021kfp}. There, it is less evident to see how causality enters the scene: how is it encoded? The question is of importance to understand more generally whether causality is a fundamental or an emergent property of space-time. 
Aspects of the causal structure of a quantum spacetime are at the roots of the spin-foam formalism \cite{Reisenberger:1996pu,Markopoulou:1997wi,Barrett:1999qw} and have been studied in \cite{Markopoulou:1997hu,Markopoulou:1999cz,Gupta:1999cp,Livine:2002rh,Pfeiffer:2002ic,Hawkins:2003vc,Oriti:2004mu,Freidel:2005bb,Oriti:2005jr,Oriti:2006wq,Livine:2006xc,Bianchi:2011hp,Rovelli:2012yy,Bianchi:2012nk,Oriti:2013aqa,Immirzi:2013rka,Cortes:2014oka,Immirzi:2016nnz,Finocchiaro:2018hks,Dona:2020yao,Jercher:2022mky}.

Our investigation proceeds as follows: In Sec.~\ref{sec:discrete_causal_structure} we recall what is meant by causality over a Lorentzian manifold and we show how it survives over a simplicial complex; in Sec.~\ref{sec:dual_skeleton} we show how the causal structure can be represented on the dual skeleton; in Sec.~\ref{sec:Lorentzian_regge_calculus} we show how the causal structure is determined by the dynamics of Regge calculus \cite{Regge:1961px}; in Sec.~\ref{sec:causal_path_integral} we elucidate the role of causality at the level of the path-integral over geometries; in Sec.~\ref{sec:BF} we investigate the case of discrete BF theory; in Sec.~\ref{sec:EPRL} we propose a causal version of the Engle-Pereira-Rovelli-Livine (EPRL) spin-foam model \cite{Engle:2007wy}; finally, in Sec.~\ref{sec:relation_to_earlier_proposals} we discuss how the proposed causal EPRL model relates to previous proposals for implementing causality such as the Livine-Oriti vertex \cite{Livine:2002rh} and the Engle vertex \cite{Engle:2011un,Engle:2015mra}.

\section{Discrete causal structure}
\label{sec:discrete_causal_structure}

The geometry of a Lorentzian manifold is fully encoded in the metric $g$.  The signature of $g$ is either $(-,+,+,+)$ or $(+,-,-,-)$. It is generally held that this freedom is a pure convention, with no physical consequences. However, for the rest of our work, it is useful to let this choice open and write the signature as $(\eta,-\eta,-\eta,-\eta)$ with $\eta \in \{-1,+1\}$.

The \textit{causal structure} of $g$ can be decomposed into two sub-notions that we call \textit{bare-causality}\footnote{The denomination is ours. Surprisingly, it seems that this notion does not carry a specific name in the literature. It is usually simply called ``causality,'' but here we need a specific name to be accurate.} and \textit{time-orientability} (see \cite{Hawking:1973a} for a standard reference).

\subsection{Bare-causality}

We call ``bare-causality'' the property of each tangent space at any point of the manifold, to be partitioned into three classes of tangent vectors: time-like, space-like and null. Formally, the classes are equivalence classes for the relation:
\begin{equation}
    u\sim v \Longleftrightarrow \text{sign} \, g(u,u) = \text{sign} \, g(v,v),
\end{equation}
where $u$ and $v$ are tangent vectors. The three equivalent classes are named:
\begin{equation}
\label{eq:def_time-like}
    \text{sign} \, g(u,u) = \left\{
    \begin{array}{c l}
        \eta  &  \textit{time-like}\\
        0 & \textit{null} \\
        - \eta & \textit{space-like}.
    \end{array} \right.
\end{equation}
The definition of bare-causality is local, in the sense that it makes sense at each point of the manifold, but it can also be formulated as a global property. The ``bare-causal structure'' of a Lorentzian space-time consists in the possibility to say, given any two points, whether they are space-like, time-like or null separated. To be clear with the definitions, two points are time-like separated if they can be joined by a smooth curve whose tangent vectors are time-like all along. It is important that the curve is smooth, because otherwise one could turn around sharply and draw a time-like curve between any two points.

\subsection{Time-orientability}

On top of bare-causality, one can define a notion of ``time-orientability.'' At a local level, time-orientability is the property of time-like vectors to be divided into two classes: past and future. Formally, the two classes are equivalence classes of time-like vectors for the relation:
\begin{equation}
\label{eq:time-orientability}
    u \sim v \Longleftrightarrow \text{sign} \, g(u,v) = \text{sign} \, g(v,v).
\end{equation}
In the case of bare-causality, the information contained in the metric $g$ alone enables to distinguish one from another the three classes of time, null and space, without ambiguity. This is not the case for time-orientability: the two classes defined by \eqref{eq:time-orientability} are perfectly symmetric. Thus, the denomination ``past'' or ``future'' is arbitrary as long as no external arrow of time is imposed additionally. So we can pick a reference vector $u_0$ (the arrow of time), label the two classes by $\omega \in \{-1,+1\}$ and say that the time-like vector $v$ is in the class $\omega$ if
\begin{equation}
\label{eq:def_omega}
    \text{sign} \, g(u_0,v) = \omega \, \eta.
\end{equation}
To fix the language, we declare that $\omega=+1$ is the future, so that $u_0$ is future-pointing.

As done previously, the local definition of time-orientation can be turned into a global one by requiring continuity across the classes at different points. Then, space-time is said to be time-orientable if it is possible to continuously define a division of time-like vectors in past and future classes. It is then possible to say that a point lies in the future of some other. As a consequence, one can define the \textit{causal future} $I^+(p)$ and the \textit{causal past} $I^-(p)$ of a point $p$. Again, the arrow of time, i.e. the labeling of ``past'' or ``future,'' is conventional, e.g., attached to a specific choice of reference vector $u_0$, but it is not a geometric property of the metric.

Time-orientability is conceptually different from bare-causality. However, any Lorentzian metric locally defines light-cones with both a bare-causal and a time-orientable structure. So the conceptual difference is often overlooked and the term of ``causality'' is used indifferently to talk about either notions or both. Yet, it is important to have the distinction clear in mind, especially when moving to quantum models, because we expect the Lorentzian metric to make way to new objects, while the underlying physical notions may survive.

\subsection{Discrete bare-causality}

At a discrete level, consider a Lorentzian 4-simplicial complex $\Delta$, i.e. a set of Minkowskian 4-simplices nicely glued together \cite{Regge:1961px}\footnote{We leave the rigorous definition of ``nicely glued together'' unspecified here as it does not affect directly our investigation. See \cite{Dona:2020yao} for details and its relation to twisted geometries \cite{Freidel:2010aq}.}. It is again possible to define the notions of bare-causality and time-orientability, both at a local and at a global level.

Each 4-simplex $\sigma$ comes with an embedding in Minkowski space-time. It is bounded by 5 tetrahedra, each of them having a unique normal $4$-vector $N$, of unit norm and directed outward. A priori, $N$ can be time-like, space-like or null. Two nearby 4-simplices share exactly one common tetrahedron $T$. The two 4-simplices are said to be time-like separated if $N$ (computed with respect to any of the two $4$-simplices) is time-like. A similar definition holds for space-like and null separation. 

Unfortunately, this local notion of bare-causality fails to extend straightforwardly to distant 4-simplices. Indeed, one could be tempted to say that two distant $4$-simplices are time-like separated if there exists a sequence of time-like separated nearby tetrahedra in-between. However, this definition fails, because, for instance, two space-like separated nearby tetrahedra could be connected by a common time-like separated nearby tetrahedron. In the previous continuous case, the smoothness of the time-like curve was preventing such a pathology, but this is not anymore possible in the discrete case. The difficulty can be circumvented by first introducing a local notion of time-orientability.

\subsection{Discrete time-orientability}

A tetrahedron is said to be space-like if it is embedded within a space-like hyperplane. In this case, its $4$-normal $N$ is time-like. Time-orientability is the property that the space-like boundary tetrahedra of a 4-simplex can be divided into two classes, by the following relation
\begin{equation}
    T_1 \sim T_2 \Longleftrightarrow \text{sign}(N_1 \cdot N_2) = \text{sign}(N_1 \cdot N_1),
\end{equation}
where the dot denotes the Minkowskian scalar product. A choice of time-orientation consists in saying which class is called past or future (relatively to the 4-simplex).

At a global level, we say that $\Delta$ is time-orientable if there exists a consistent choice of time-orientation for each 4-simplex, so that each space-like tetrahedron has an opposite time-orientation relatively to each of the two 4-simplices that bounds it: if a tetrahedron is in the future of a 4-simplex, it should be in the past of another.

Given two 4-simplices, $\sigma_1$ and $\sigma_2$, sharing a space-like tetrahedron $T$, we say that $\sigma_2$ is in the future of $\sigma_1$ if $T$ is in the future of $\sigma_1$ (hence in the past of $\sigma_2$). This definition allows us to define straightforwardly a notion of causal future and causal past of a 4-simplex: $\sigma_2$ is in the future of $\sigma_1$ if there exists a future-oriented chain of 4-simplices in-between. This definition encompasses the notion of time-like separation for distant 4-simplices that was initially looked for. Thus, both bare-causality and time-orientability are defined locally and globally in the discrete setting.

\section{Causality on the dual skeleton}
\label{sec:dual_skeleton}

The previously defined discrete causal structure can be easily represented on $\Delta^*_1$, the dual 1-skeleton of $\Delta$. The  1-skeleton $\Delta^*_1$ is built from $\Delta$ by replacing each $4$-simplex by a vertex, each tetrahedron by an edge and forgetting about triangles, segments and points of $\Delta$.

Bare-causality discriminates between space-like and time-like edges\footnote{We deliberately ignore the null case, which does not seem to shed much light on our investigation.}, while time-orientability provides an orientation to the time-like edges. Overall, causality is then represented by
\begin{enumerate}
    \item An arrow from past to future on time-like edges,
    \item No arrow on space-like edges.
\end{enumerate}
In the following, we assume that all tetrahedra are space-like, which implies that all $N$ are time-like. Dually, it means that all the edges of $\Delta_1^*$ carry an arrow. This simplifying assumption is made in many of the formulations of spin-foams. It is important to note that this condition automatically implements some implicit assumptions about the fundamental causal structure of space-time. Indeed, this assumption erases any local notion of bare-causality: all nearby 4-simplices are time-like separated. Thus, at the most local level, the primacy is granted to time-orientability. The notion of bare-causality only emerges at a more global level as follows: given two distant vertices, if one is not in the past of the other, then they are said to be space-like separated.

\subsection{Dual causal set}

In mathematical terms, $\Delta_1^*$ is a 5-valent \textit{simple oriented graph}\footnote{A \textit{directed graph} is given by a set of vertices and a set of ordered pairs of vertices (arrows). It is said \textit{simple} if there are no arrows from a vertex to itself. It is said \textit{oriented} if there is at most one arrow between any two vertices.}. Assuming it is also \textit{acyclic}\footnote{A directed graph is \textit{acyclic} if it has no directed cycles, which means, in causal language, no closed time-like curves. If we do assume the presence of directed cycles, then the construction of the poset is still possible but subtler because the anti-symmetry implies the contraction of such cycles, so that more combinatorial information is lost.}, its \textit{transitive closure} defines a \textit{poset} (partially ordered set). The elements of the poset are the vertices of $\Delta_1^*$ and the partial order $\leq$ comes from a unique extension of the set of arrows with the following properties:
\begin{enumerate}[itemsep=0mm]
    \item Reflexivity: $v \leq v$ (by convention).
    \item Anti-symmetry: $v_1 \leq v_2$ and $v_2 \leq v_1$ imply $v_1 = v_2$. 
    \item Transitivity: $v_1 \leq v_2$ and $v_2 \leq v_3$ imply $v_1 \leq v_3$.
\end{enumerate}
In most reasonable cases, the poset of $\Delta_1^*$ is \textit{locally finite}, meaning that for any pair of vertices $(v_1,v_2)$, the so-called \textit{causal diamond} $\left\{ v \mid v_1 \leq v \leq v_2 \right\}$ is a finite set. Such a poset is a \textit{causal set}, as defined originally in \cite{Bombelli:1987aa}.

We have shown, without much surprise, that the discretization of a Lorentzian manifold naturally carries a causal set structure. Causal set theory takes the causal set structure as a starting point. Then, the question naturally poses itself as to whether or not it is possible to reconstruct $\Delta^*_1$ from its associated causal set only. Given a causal set, one can derive a notion of neighborhood by declaring that two vertices $x$ and $y$, such that $x \leq y$, are next to each other if there is no $z \neq x,y$ such that $x \leq z \leq y$. In other words, the neighborhood relations are obtained by a \textit{transitive reduction} of the causal set, i.e., a graph with the fewest possible arrows and the same ``reachability relations'' as the causal set. Interestingly, for a finite directed acyclic graph, such a transitive reduction is unique. However, \textit{the transitive reduction of the transitive closure is not the identity}. Thus, it is not possible to recover $\Delta^*_1$ from the causal set by transitive reduction. In other words, the notions of neighborhood for causal set theory and for discrete Lorentzian geometry, as described over $\Delta^*_1$, are not the same.

Similarly, the conformal factor, which is an important piece of information of the metric, can arise in several different ways. In causal set theory, it emerges by counting the number of vertices within a given causal diamond \cite{Bombelli:1987aa,Surya:2019ndm}. In discrete Lorentzian geometry, it can be given by the Lorentzian volume of the 4-simplices, which requires additional input, not deducible from the causal set alone. For instance, the additional information can be provided by coloring each vertex with a real number (the 4-simplex volume), or by coloring each edge with the volume of the corresponding tetrahedron, or by introducing faces and coloring them with the area of the corresponding triangles. The latter option is of course relevant for spin-foam models as discussed also in \cite{Livine:2002rh,Cortes:2014oka,Wieland:2014nka,Immirzi:2016nnz}.

To work algebraically with the causal structure, it will soon appear convenient to express the orientation of the edges as follows. Given a vertex $v$, we define the orientation of an edge\footnote{We denote indifferently $e \in v$ or $v \in e$ when the vertex $v$ is an endpoint of the edge $e$.} $e \in v$ with respect to $v$ as
\begin{equation}
\label{eq:edge_orientation}
    \varepsilon_v(e) \overset{\text{def}}= \left\{
    \begin{aligned}
        - 1 \quad &\text{if $e$ is incoming}, \\
        1 \quad &\text{if $e$ is outgoing}.
    \end{aligned}
    \right.
\end{equation}
This convention is similar to the earlier choice in equation \eqref{eq:def_omega} to call $\omega=1$ the future.
We define a \textit{causal structure on the 1-skeleton} as an assignment of an orientation $\varepsilon_v(e)$ to each pair $(v,e)$ such that $e\in v$, under the constraint\footnote{In equation \eqref{eq:edge_orientation}, a global convention was chosen to attribute a numerical value to `incoming' and `outgoing'. This convention can be made local by introducing at each vertex $v$ a variable $\mu_v \in \{-1,1\}$, and defining instead
\begin{equation}
    \varepsilon_v(e) \overset{\text{def}}= \left\{
    \begin{aligned}
        - \mu_v \quad &\text{if $e$ is incoming} \\
        \mu_v \quad &\text{if $e$ is outgoing}.
    \end{aligned}
    \right.
\end{equation}
In this case, the constraint becomes
\begin{equation}
    \mu_{v_1}  \varepsilon_{v_1}(e) = - \mu_{v_2} \varepsilon_{v_2}(e),
\end{equation}
which is more similar to what can be found in \cite{Livine:2002rh}.}
\begin{equation}
\label{eq:gluing_edges}
    \varepsilon_{v_1}(e) = -  \varepsilon_{v_2}(e),
\end{equation}
where $v_1$ and $v_2$ are the two end-points of $e$. The latter condition expresses the fact that an incoming edge, with respect to one vertex, is outgoing with respect to the other.

\subsection{Causal wedges}

\begin{figure}[t]
    \centering
        \begin{overpic}[width=0.2 \textwidth]{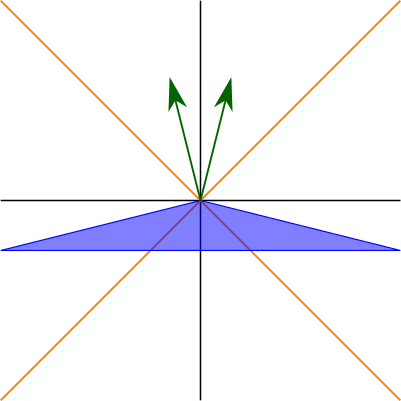}
    \put (55,85) {$N_2$}
    \put (35,85) {$N_1$}
    \end{overpic}
    \hspace{8em}   
     \begin{overpic}[width=0.2 \textwidth]{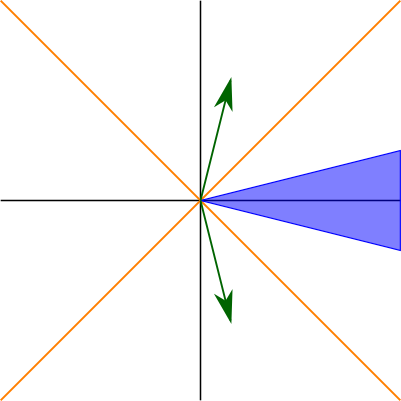}
    \put (55,85) {$N_1$}
    \put (55,10) {$N_2$}
    \end{overpic}
\caption{Left: thick wedge, co-chronal normals. Right: thin wedge, anti-chronal normals.}
\label{fig:wedge}
\end{figure}

We have seen that the causality of $\Delta$ can be read on the edges of $\Delta^*_1$. Now we are going to show that it can also be read equivalently on the wedges of the dual 2-skeleton $\Delta^*_2$.
To proceed, let's go back to the 4-simplicial complex $\Delta$. A pair $w=(t,\sigma)$ such that $t \in \sigma$ is called a \textit{wedge}. Given a wedge $w$, there exists exactly two tetrahedra $T_1,T_2 \in \sigma$ that share $t$, to which are associated the normals $N_1$ and $N_2$. The \textit{dihedral angle} of $w$ is defined as
\begin{equation}
\label{eq:dihedral_angle}
    \theta_w \overset{\text{def}}= \text{sign}(N_1 \cdot N_2 ) \cosh^{-1}(|N_1 \cdot N_2 |).
\end{equation}
This definition is a natural extension of the notion of dihedral angle from Euclidean to Minkowskian geometry. Its absolute value $|\theta_w|$ depends only on the absolute value  of the scalar product $|N_1 \cdot N_2 |$ of the normals. On the other hand, its sign depends on if the relative time-orientation of the two normals
\begin{equation}
  \text{sign}(\theta_w)= \left\{ 
    \begin{aligned}
        +\eta \quad &\text{if $N_1$ and $N_2$ are co-chronal} \\
        -\eta \quad &\text{if $N_1$ and $N_2$ are anti-chronal}.
    \end{aligned}
    \right.
\end{equation}
When the normals are co-chronal (resp. anti-chronal), the wedge is said to be \textit{thick} (resp. \textit{thin}). At the level of the wedges, causality shows up as follows: a thick wedge encloses a time-like region, while a thin wedge encloses a space-like region (see fig. \ref{fig:wedge}). The thin/thick distinction provides an orientation of the wedges. However this orientation does not extend to triangles because several wedges of the same triangle may not have the same orientation.

This notion translates easily on the dual complex. A pair of a face and a vertex $(f,v)$ with $f\in v$ defines a (dual) wedge on the 2-skeleton $\Delta^*_2$. There exists two unique edges $e_1$ and $e_2$ such that $e_1, e_2 \in f$ and $e_1, e_2 \in v$. The wedge is \textit{thick} if $e_1$ and $e_2$ are both incoming or both outgoing. It is \textit{thin} otherwise.
Algebraically, the wedge orientation can be defined as
\begin{equation}
    \varepsilon_v(f) \overset{\text{def}}= \left\{ 
    \begin{aligned}
        +\eta \quad &\text{if thick}, \\
        -\eta \quad &\text{if thin}.
    \end{aligned}
    \right.
\end{equation}
It is then easy to show that
\begin{equation}
\label{eq:epsilon_f_from_e}
    \varepsilon_v(f) = \eta \, \varepsilon_v(e_1) \varepsilon_v(e_2).
\end{equation}

As we have presented it, the wedge orientation is a byproduct of the edge orientation. However, one can wonder whether it is possible to go the other way around and to compute the $\varepsilon_v(e)$ as  a function of the $\varepsilon_v(f)$, i.e. to invert equation \eqref{eq:epsilon_f_from_e}? The short answer is \textit{no}, but not much information is actually missing to do this inversion.

\begin{figure}[t]
    \centering
         \includegraphics[width=0.2 \textwidth]{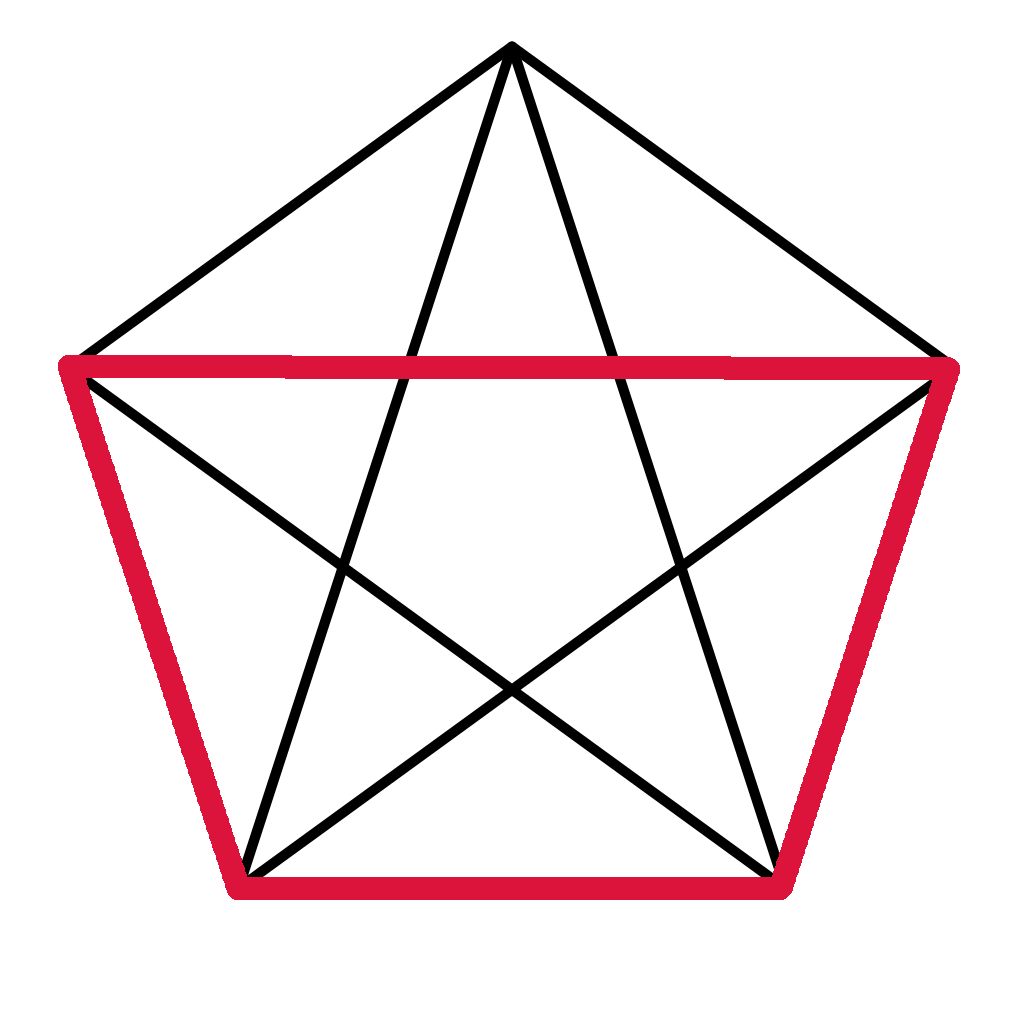}
         \hspace{1em}
        \includegraphics[width=0.2 \textwidth]{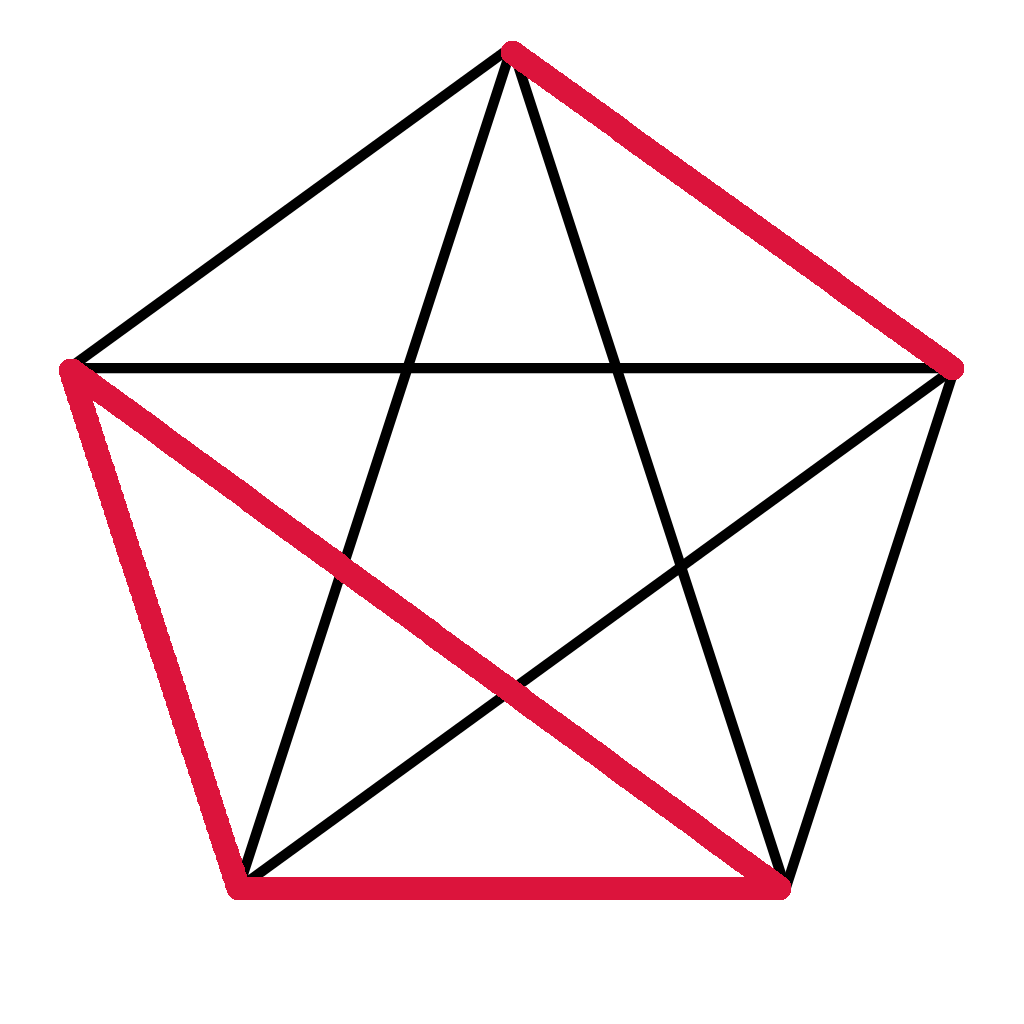}
         \hspace{4em}
        \includegraphics[width=0.2 \textwidth]{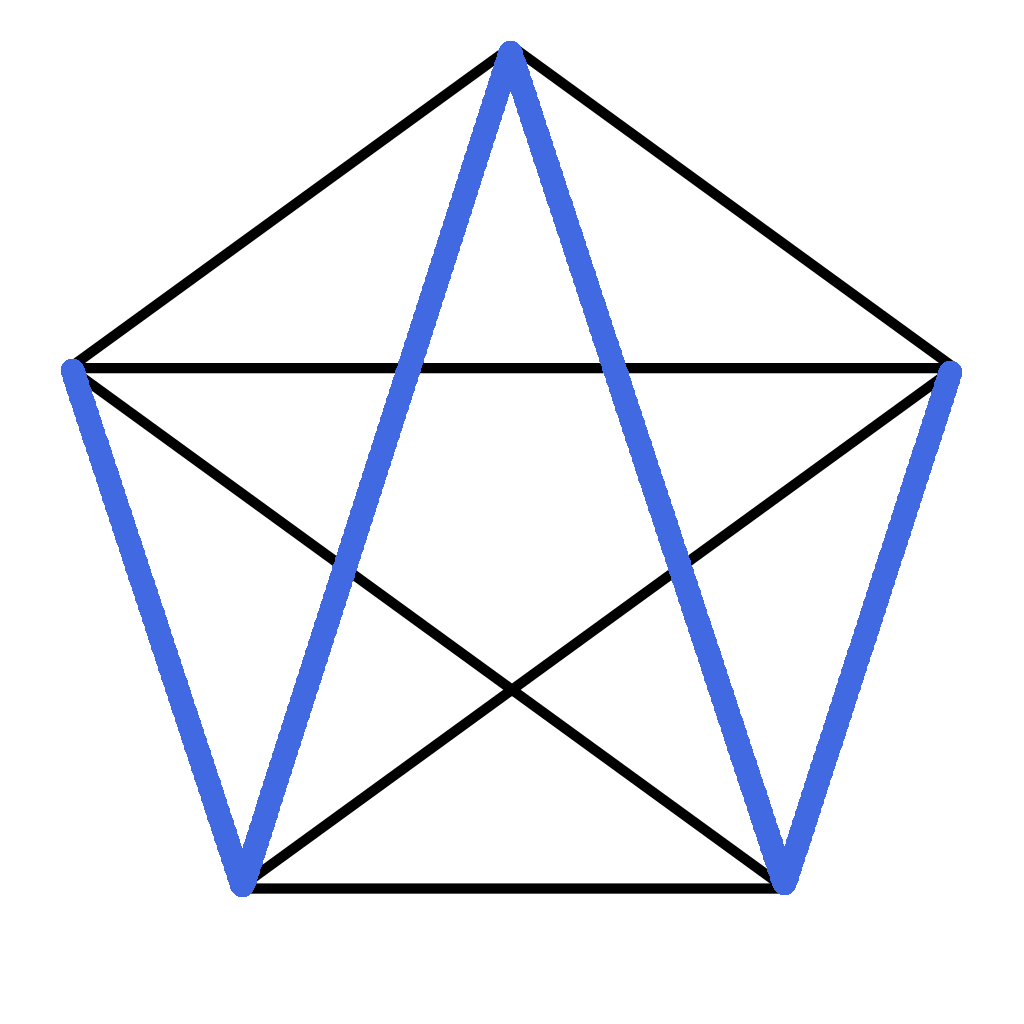}
        \hspace{1em}
        \includegraphics[width=0.2 \textwidth]{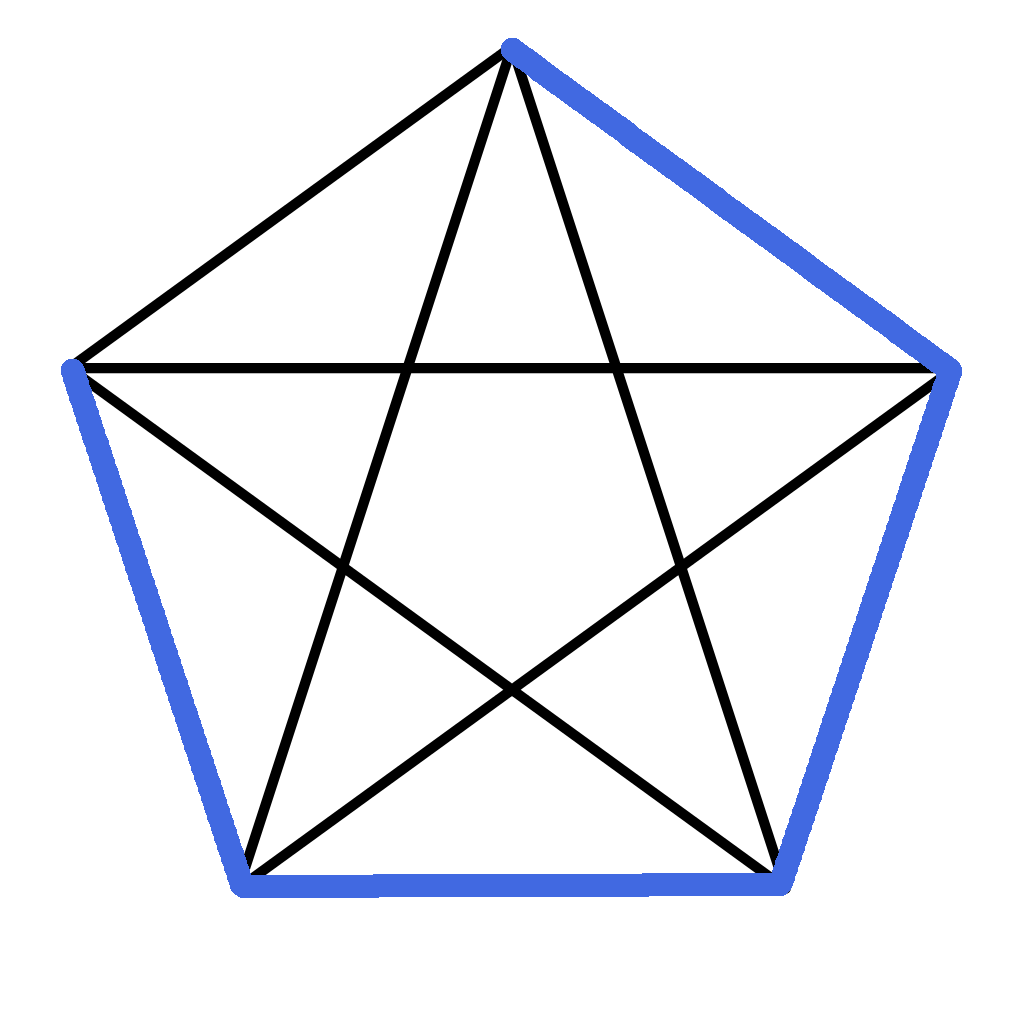}
\caption{Left (Red): Two examples of sets of 4 wedges which form cycles. Right (Blues): Two examples of sets of 4 wedges which do \textit{not} form cycles.}
\label{fig:pentagram_in_pentagon}
\end{figure}

Around the same vertex $v$, equation \eqref{eq:epsilon_f_from_e} defines a system of 10 equations (one per face) with 5 unknowns (one per edge), so we may fear it to be over-constrained. However, the rank of the system \eqref{eq:epsilon_f_from_e} is only 4. Indeed, given the orientation $\varepsilon_v(f)$ for any 4 wedges \textit{that do not form a cycle}, one can deduce the orientation of the other 6. A cycle is a sequence of faces that closes, bounding a 3D angle. Rather than a formal definition, this notion is best understood through few examples on the links of the vertex graph\footnote{Given a vertex, the vertex graph associates a node to each edge and a link to each wedge in-between. In graph theory, the word ``edge'' is usually used instead of ``link''. But we stick to a common convention in loop quantum gravity (see \cite{Rovelli:2014a}) where ``edge'' is reserved to the bulk of 2-complexes and ``link'' is used for the boundary.} (see figure \ref{fig:pentagram_in_pentagon}).
Then it is easy to show that the product of orientations along any cycle of wedges is
\begin{equation}\label{eq:cycle}
    \prod_{f \in \text{cycle}} \varepsilon_v(f) = \eta^{\# \text{cycle}},
\end{equation}
where $\# \text{cycle}$ is the number of faces $f$ in the cycle. Since any set of 5 wedges around $v$ contains a cycle, then $4$ wedges orientation are indeed sufficient to fix them all. Hence the system \eqref{eq:epsilon_f_from_e} is actually under-constrained of exactly one dimension.

Given a vertex $v$, denote the surrounding edges by $e_i$, with $i \in \{1,...,5\}$, and accordingly the surrounding faces by $f_{ij}$. To make the system invertible, let us add one independent equation, by defining the orientation of the vertex $v$ as:
\begin{equation}
\label{eq:definition_e_v}
    \varepsilon_v \overset{\text{def}}=  \prod_j \varepsilon_v(e_j) .
\end{equation}
Then, the set of equations \eqref{eq:epsilon_f_from_e} augmented of \eqref{eq:definition_e_v}, with unknowns $\varepsilon_v(e)$, is invertible and one can show that 
\begin{equation}
    \varepsilon_v(e_i) = \varepsilon_v \prod_{k \neq i} \varepsilon_v(f_{ik}).
\end{equation}
So we see that we can recover the orientation of the edges from the orientation of the wedges, up to a vertex orientation.

Let's now consider a skeleton with many vertices. Can we deduce the orientation of the edges of the 1-skeleton from the orientation on the wedges of the 2-skeleton? From the previous analysis with a single vertex, we know that it will be possible to invert the system of equations if one considers in addition one orientation $\varepsilon_v$ per vertex. However, the set of $\varepsilon_v$ is itself constrained, because of the gluing condition \eqref{eq:gluing_edges}, which now reads 
\begin{equation}
    \varepsilon_{v_1} \prod_{f | e \in f} \varepsilon_{v_1}(f) = - \varepsilon_{v_2} \prod_{f | e \in f} \varepsilon_{v_2}(f)\,.
\end{equation}
This constraint eliminates almost all the degrees of freedom introduced by $\varepsilon_v$, so that there remains finally only the freedom to fix the orientation of a single vertex in the whole skeleton. The orientation of all others can be deduced from it and the $\varepsilon_v(f)$. Indeed, assume that you have a 2-skeleton where all the $\varepsilon_v(f)$ have been fixed (satisfying the constraint along cycles). Then, if you fix only one $\varepsilon_v$, the orientation of the edges around $v$ are fixed, and this ``orientation-fixing'' will then propagate everywhere else, so the full set of $\varepsilon_v$ will ultimately be fixed. Reversing the orientation of one $\varepsilon_v$, will reverse the entire skeleton, which corresponds to the time-reversal symmetry.

We have previously shown that the causal structure of discrete general relativity can be encoded in the dual 1-skeleton with oriented edges. Now we have just seen that the \textit{causal structure of a 2-skeleton} can be described as the assignation of $\varepsilon_v(f)$ to each wedge under the cycle constraint \eqref{eq:cycle} and a global orientation $\varepsilon$, which can be regarded as a global arrow of time.

\section{Lorentzian Regge calculus}
\label{sec:Lorentzian_regge_calculus}
So far, we only focused on the kinematical aspects of causality. Now, we will see how causality shows up in the dynamics \cite{Regge:1961px}.  
\subsection{Lorentzian Regge action}
Following \cite{Livine:2002rh}, we extend the formulae of \cite{Barrett:1993db} to the $4$-dimensional case, the Lorentzian Regge action is a sum over the triangles $t$:
\begin{equation}
    S_R \overset{\text{def}}= \sum_t A_t \delta_t,
\end{equation}
with $A_t$ the area of the triangle $t$, and $\delta_t$ the deficit angle defined as a sum over the 4-simplices $\sigma$ surrounding $t$:
\begin{equation}
    \delta_t \overset{\text{def}}= \sum_{\sigma | t \in \sigma} \theta_{t\sigma},
\end{equation}
with the dihedral angle defined by equation \eqref{eq:dihedral_angle}. The order of the two sums can be exchanged:
\begin{equation}
    S_R = \sum_\sigma \sum_{t | t \in \sigma} A_t \theta_{t\sigma}.
\end{equation}
To derive the equations of motion by variational calculus, one should tell which are the independent variables of which $S_R$ is a function of. In the original Regge calculus, it is shown that if the action is considered as a function of the lengths $l_s$ of the segments $s$, the resulting equations of motions become Einstein equations in the continuous limit. However, this is not the only possible choice.

\subsection{First-order Regge calculus}

Barrett has proposed a formulation where the independent variables are both the lengths $l_s$ and the angles $\theta_{t\sigma}$ \cite{Barrett:1994nn}. This choice of variables mimics the Palatini formulation which takes the metric and the torsion-less connection as primary fields of the Einstein-Hilbert action. Compared to the original Regge calculus, the introduction of $\theta_{t\sigma}$ extends the total number of variables. In order to recover the equations of motion, it is then necessary to add constraints to the action, which is done with a Lagrange multiplier $\mu_\sigma$ per each 4-simplex $\sigma$. One obtains
\begin{equation}
    S[l_s,\theta_{t\sigma}, \mu_\sigma]  = \sum_\sigma \sum_{t | t \in \sigma} A_t(l_s) \theta_{t\sigma} + \sum_\sigma \mu_\sigma \det \gamma_\sigma
\end{equation}
where $\gamma_\sigma$ is the $5 \times 5$ matrix whose elements are the Minkowskian scalar products between the normals to the boundary tetrahedra:
\begin{equation}
    [\gamma_\sigma]_{ij} = N_i \cdot N_j = \text{sign}(\theta_{ij}) \cosh \theta_{ij},
\end{equation}
where $N_i$ is the (unit outward) normal to the $i$th boundary tetrahedron of $\sigma$ and $\theta_{ij}$ is the dihedral angle between the tetrahedra $i$ and $j$. The Lagrange multiplier imposes the constraint
\begin{equation}
    \det \gamma_\sigma = 0
\end{equation}
which implements the closure of the normals, i.e.
\begin{equation}
    \sum_i V_i N_i = 0,
\end{equation}
with $V_i$ the volume of the $i$th tetrahedron.

\subsection{Causal structure from dynamics}

This choice of variables for the action makes it clear how causality plays its role in the dynamics. Indeed, as explained previously, the causal structure is encoded on the wedges as the sign of the dihedral angle between two neighbouring tetrahedra. This information can be directly obtained from the sign of the variable $\theta_{t\sigma}$. However, any configuration of signs of $\theta_{t\sigma}$ does not define an allowed causal structure as it must also satisfy the cycle constraint \eqref{eq:cycle}. So, an assignment of $\theta_{t \sigma}$ defines an \textit{orientation structure}, but not necessarily a \textit{causal structure}.

Meanwhile, it is also possible to obtain causal information from the lengths $l_s$. First of all, the sign of $l_s$ tells whether the segment is space-like or time-like. Here we are assuming that the segments are all space-like. Nevertheless, it is still possible to extract additional causal information. If some geometrical constraints are satisfied (generalisation of the triangular inequalities), then it is possible to reconstruct uniquely the geometry of a 4-simplex from the lengths of its segments. Then, there exists a formula expressing the dihedral angle as a function of the lengths $l_s$. This second derivation of the dihedral angle does not necessarily match with $\theta_{t \sigma}$ because $\theta_{t \sigma}$ and $l_s$ are taken as independent variables in the first-order Regge calculus. So, the variables $\theta_{t \sigma}$ and $l_s$ define independently two coexisting notions of causal structure on the wedges. 

Of course, the two notions of causal structure must match when the equations of motion are imposed. In particular, the cycle constraint appears as a corollary of the equations of motion. Indeed the constraint $\det \gamma_\sigma = 0$ implies the existence of a vector $v$ such that $\sum_i \gamma_{ij} v_i = 0$. Then the equation of motion obtained by varying $\theta_{ij}$ yields
\begin{equation}
    A_t = \kappa_\sigma \mu_\sigma v_i v_j \sinh \theta_{ij}
\end{equation}
for some $\kappa_\sigma \in \mathbb{R}$ (see \cite{Barrett:1994nn}). Since $A_t > 0$, 
\begin{equation}
    \text{sign} \, \theta_{ij} = \text{sign}(\kappa_\sigma \mu_\sigma) \, \text{sign} \, v_i \, \text{sign} \, v_j 
\end{equation}
which implies the constraints \eqref{eq:cycle}. This shows that the equations of motion impose the structure of the wedges to be causal. This result should be stressed. In the standard metric formulation of general relativity, the causal structure is already well-defined at the level of the kinematics, because any metric defines a causal structure. In the first-order Regge calculus, we see that the causal structure does not necessarily exist at the kinematical level, but the equations of motion impose the cycle constraint, which selects a surface of variables where every configuration has a well-defined causal structure.

\section{Causal path integral}
\label{sec:causal_path_integral}

So far, the analysis was purely classical, although discrete. When going to the quantum regime of gravity, it is reasonable to expect that the generic state of the metric field is a superposition of classical configurations. In particular, there might not be a definite causal structure. Several causal histories may interfere and thus generate in principle observable effects. A theory of quantum gravity should be able to predict these effects through the computation of transition amplitudes between different states of space.

Let's clarify the main ideas by proceeding heuristically, although a precise mathematical formulation may be more difficult to achieve. The standard procedure starts by foliating the space-time manifold into constant-time slices: $\mathcal{M} \cong \Sigma \times \mathbb{R}$. The classical states are 3-metrics $h$ defined over $\Sigma$. At the quantum level, the 3-metric is an operator $\hat h$, with eigenstates $\ket{h}$ whose eigenvalues are the classical 3-metrics $h$. The sum-over-histories approach to quantum gravity \cite{Misner:1957wq,Wheeler:1964qna,Hawking:1978jz} proposes to compute the transition amplitude between the state $\ket{h_0}$ at time $t_0$ and the state $\ket{h_1}$ at time $t_1$ as a path integral:
\begin{equation}\label{eq:standard_transition}
    \braket{h_1}{h_0} = \int [\mathcal{D}g] \, e^{\frac{i}{\hbar} S[g]}\,,
\end{equation}
where $[\mathcal{D}g]$ is a measure on the set of 4-metrics $g$ over $\Sigma \times [t_0,t_1]$, such that the restriction of $g$ to the slice $t=t_0$ (resp. $t=t_1$) is $h_0$ (resp. $h_1$), and $S[g]$ is the Einstein-Hilbert action evaluated on such a metric.

\subsection{General boundary formulation}

The previous and standard formulation is not ideal because it relies upon a slicing of space-time into constant-time leaves, which may already fix too much structure for a general treatment of causality. Better suited for our purpose is the general boundary formulation developed by Oeckl in \cite{Oeckl:2005bv}. Consider a region of space-time $M$ with boundary $\Sigma$. A 4-metric $g$ on $M$ induces a 3-metric $h$ on $\Sigma$. The crux of a quantum theory of space-time is the computation of the \textit{metric propagator}:
\begin{equation}\label{eq:metric_propagator}
    Z_M(h) = \int [\mathcal{D}g] \, e^{\frac{i}{\hbar} S[g] }
\end{equation}
where the integral is carried over all the 4-metrics $g$ bounded by $h$. The standard formulation (equation \eqref{eq:standard_transition}) is recovered when $\Sigma$ is made of two disconnected components (past and future). In general, the boundary will have space-like, time-like and possibly null components. Here we restrict attention to a finite space-time region with boundary consisting of two space-like components. This lense-shaped space-time region is foliated by finite space-like leaves which meet at a fixed $2$-dimensional corner.

\subsection{Regge path integral}

As a way towards the actual computation of the metric propagator, one can discretise the previous formula. Consider a 4-simplicial complex $\Delta$. Its boundary is a 3-simplicial complex $\Sigma$. Working with the Regge action, the metric propagator is a function of the length of the segments of $\Sigma$, and it reads:
\begin{equation}\label{eq:regge_propagator}
    \mathcal{A}_\Delta(l_\Sigma) = \int [\dd l_s] \, e^{\frac{i}{\hbar} S_R[l_s]}.
\end{equation}
The integral is done over the lengths $l_s$ of all the segments $s$ in the bulk of $\Delta$. Note that each integral could also be replaced by a sum with a cut-off, in order to ensure a finite value to the propagator, but it does not seem useful in our quest for causality.

Using the first-order Regge calculus, the propagator reads: 
\begin{equation}
    \mathcal{A}_\Delta(l_\Sigma,\theta_\Sigma,\mu_\Sigma) = \int [\dd l_s] [\dd \theta_{t\sigma}] [\dd \mu_\sigma] 
     \prod_\sigma e^{\frac{i}{\hbar} \left( \sum_{t | t \in \sigma} A_t \theta_{t\sigma} +  \mu_\sigma \det \gamma_\sigma \right) }\,.
\end{equation}
The integration over $\mu_\sigma$ can be formally carried over, which gives a $\delta$-function that fixes the constraint:
\begin{equation}\label{eq:first_order_regge_propagator}
    \mathcal{A}_\Delta(l_\Sigma,\theta_\Sigma) = \int [\dd l_s] [\dd \theta_{t\sigma}] \prod_\sigma \delta(\det \gamma_\sigma) \, e^{\frac{i}{\hbar} \sum_{t | t \in \sigma} A_t \theta_{t\sigma}  }\,.
\end{equation}

\begin{figure}[t]
\hspace{4em}
\begin{overpic}[width=0.3 \textwidth]{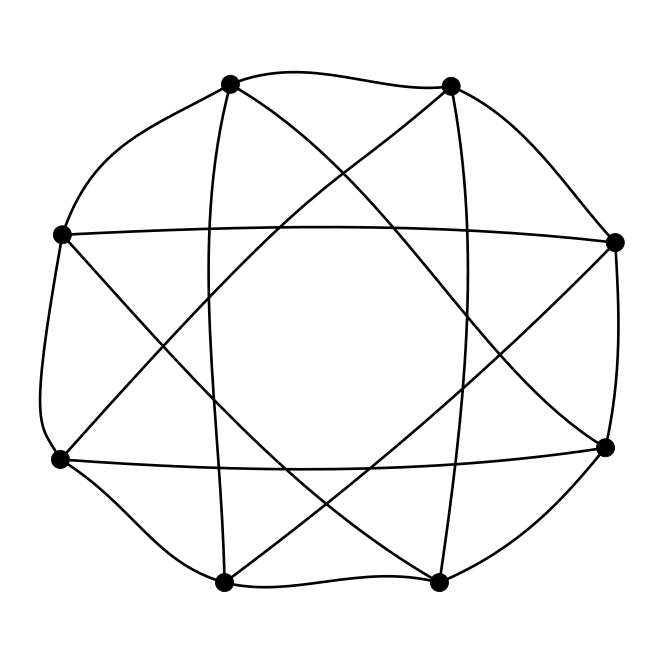}
    \put (30,90) {\textbf{+}}
    \put (70,90) {\textbf{+}}
    \put (0,65) {\textbf{--}}
    \put (95,65) {\textbf{+}}
    \put (0,30) {\textbf{+}}
    \put (95,30) {\textbf{+}}
    \put (25,10) {\textbf{--}}
    \put (70,10) {\textbf{--}}
\end{overpic}
\hspace{4em}
\begin{overpic}[width=0.3 \textwidth]{gfx/boundary_causality.png}
    \put (50,90) {\textbf{+}}
    \put (10,80) {\textbf{--}}
    \put (0,50) {\textbf{--}}
    \put (15,15) {\textbf{--}}
    \put (50,10) {\textbf{+}}
    \put (85,20) {\textbf{--}}
    \put (95,50) {\textbf{+}}
    \put (85,75) {\textbf{+}}
        \put (47,63) {\textbf{--}}
    \put (35,55) {\textbf{+}}
    \put (35,45) {\textbf{--}}
    \put (37,37) {\textbf{+}}
    \put (46,32) {\textbf{+}}
    \put (60,38) {\textbf{--}}
    \put (65,45) {\textbf{--}}
    \put (60,55) {\textbf{+}}
\end{overpic}
\caption{Example of causal structure on a boundary graph. Left: encoded on the nodes. Right: encoded on the links.}
\label{fig:boundary_causality}
\end{figure}

\subsection{Causal structure of the boundary}

A causal structure on $\Delta$ induces a causal structure on its boundary $\Sigma$. It consists in saying for each tetrahedron of the boundary whether it shall be regarded as future or past. It can be represented on the boundary of the dual 2-skeleton $\Delta^*_2$, which is a 4-valent graph. The induced causal structure consists in assigning a sign to each node, depending on whether the edge attached to it is pointing inside or outside the bulk. Conventionally, we take the sign to be positive for an outward edge (future tetrahedron) and negative for inward edge (past tetrahedron). An example is shown in figure \ref{fig:boundary_causality}. It is important to notice that fixing a causal structure on the boundary does not in general impose a single causal history in the bulk: many different histories may share the same causal boundary.

The causal information of the boundary can also be encoded on the links. To each link, one associates the product of the sign of the endpoints (see fig. \ref{fig:boundary_causality}). For a 4-valent graph, there are twice as many links as there are nodes. But despite the double number of variables, this encoding is not injective, but 2-to-1. Physically, by encoding causality on the links, we only provide information about bare-causality, while encoding over the nodes also gives a time-orientation.

A random assignment of signs to links only defines a causal structure on the boundary if it enables to consistently assign signs to the nodes. This happens if, and only if, the signs on the links satisfy the constraint that their product around any loop of the graph is $1$. The latter constraint is implied by the cycle condition \eqref{eq:cycle} in the bulk. In fact, when two edges crossing the boundary share a common vertex, the sign of the wedge matches the sign of the link between the two corresponding nodes. More generally, the sign of the link is equal to the product of the signs of the wedges around the corresponding face in the bulk, which can be written
\begin{equation}\label{eq:boundary_bulk_relation}
    \varepsilon_l = \prod_{v \in f} \varepsilon_v(f),
\end{equation}
where $f$ is the face that intersect the boundary along the link $l$.

\subsection{Causal amplitude}

The metric propagator is a function of the boundary variables. In concrete situations, the bare-causal structure of the boundary may be fixed by an assignment of links orientations $\varepsilon_l$. In this case, the range of integration on the angles $\theta_{t\sigma}$ in equation \eqref{eq:first_order_regge_propagator} must be restricted for the wedges that belong to faces intersecting the boundary. This restriction consists in implementing the constraint \eqref{eq:boundary_bulk_relation}. For instance, consider a link $l$ bounding a face $f$ (dual to $t$) with only one vertex $v$ (dual to $\sigma$). If $\varepsilon_l = 1$ (resp. $\varepsilon_l = -1$), then the integration over $\theta_{t\sigma}$ shall be carried over $\mathbb{R}^+$ (resp. $\mathbb{R}^-$), instead of $\mathbb{R}$.  

For the $\theta_{t\sigma}$ in the bulk, the integration is still done over all $\mathbb{R}$. We can rewrite the amplitude \eqref{eq:first_order_regge_propagator} as
\begin{equation}
\label{eq:causal_amplitude}
    \mathcal{A}_\Delta(l_\Sigma, r_\Sigma,\varepsilon_\Sigma) = \sum_{[\varepsilon_{t\sigma}]} \int [\dd l_s] [\dd r_{t\sigma}] \prod_\sigma \delta(\det \gamma_\sigma) \, e^{\frac{i}{\hbar} \sum_{t | t \in \sigma} A_t \varepsilon_{t\sigma} r_{t\sigma}}\,,
\end{equation}
where the sum is done over all possible orientations $\varepsilon_{t\sigma}$ of wedges in the bulk, compatible with the bare-causality of the boundary $\varepsilon_\Sigma$, and the integration in $r$ is performed over $\mathbb{R}^+$. Thus, the path-integral is summing over both configurations which satisfy and do not satisfy the cycle condition \eqref{eq:cycle}. 

To be more precise, at each vertex, the surrounding wedge orientations can satisfy the cycle condition either for $\eta=1$, or for $\eta=-1$, or none of them. The two first cases correspond to the possibility of locally defining a light-cone. These light-cone are only local because the value of the signature $\eta$ may disagree from one 4-simplex to another. There is a global causal structure only when the signature is the same for all 4-simplices. So the sum over orientations can be decomposed into several terms:
\begin{equation}
    \sum_{[\varepsilon_{t\sigma}]} = \sum_{\substack{[\varepsilon_{t\sigma}] \\ \text{causal} \ \eta=1}} + \sum_{\substack{[\varepsilon_{t\sigma}] \\ \text{causal} \ \eta=-1}} + \sum_{\substack{[\eta_\sigma] \\ \text{signature} \\ \text{changes}}} \sum_{\substack{[\varepsilon_{t\sigma}] \\ \text{local} \\ \text{light-cones}}} + \sum_{\substack{[\varepsilon_{t\sigma}] \\ \text{spurious}}}
\end{equation}
The latter terms gather ``non-causal'' or ``spurious'' configurations in the sense that there is at least one 4-simplex for which the set of wedge orientations $\varepsilon_{t \sigma}$ do not define a consistent local light-cone. The presence of such non-causal histories contributing to the amplitude is the consequence of the peculiar choice of variables of the first-order Regge calculus, using both the lengths $l_s$ and the dihedral angles $\theta_{t \sigma}$. In the standard Regge calculus, only the lengths $l_s$ are used, and thus the path-integral is only summing over causal configurations.

However, in the classical limit, when $\hbar \rightarrow 0$, the configurations that contribute the most are the stationary points of the action, which satisfy the classical equations of motion and thus, as seen previously, have a proper causal structure. So the causal configurations are selected in the classical limit. These configurations can be partitioned in two subsets depending on the choice of signature $\eta \in \{-1,1\}$ for which the cycle condition \eqref{eq:cycle} is satisfied.

In this framework, the existence of a consistent causal structure is an emerging feature of space-time. It originates from orientation degrees of freedom located on the wedges. At the quantum level, only a minority of configurations define proper light-cones. The unveiling of this structure offers the possibility to define alternative amplitudes by fixing some of the degrees of freedom.

For instance, one could keep in \eqref{eq:causal_amplitude} only the terms $[\varepsilon_{t\sigma}]$ which satisfy the cycle condition \eqref{eq:cycle}. One could also further restrict the sum by imposing a choice of signature $\eta$. Eventually, one could keep only a single term that has a fixed causal structure.

The suggestion of considering such restrictions in the range of the path-integral was initially suggested by Teitelboim \cite{Teitelboim:1981ua}  and first applied to spinfoams by Livine and Oriti \cite{Livine:2002rh}. In the context of the standard formulation (equation \eqref{eq:standard_transition}) $h_0$ is regarded as ``past'' and $h_1$ as ``future''. So Teitelboim proposed to restrict the range of integration over the 4-metrics $g$ for which the proper time from the first to the second slice is positive. Working in the ADM formalism, it amounts to restricting the range of integration of the lapse $N$ to positive values only. The resulting ``causal amplitude'' is not anymore gauge-invariant, i.e. it is not a solution of the Hamiltonian constraint.

However, Teitelboim argued that such a restriction could be worth considering by drawing an analogy with the propagator of the free relativistic particle. Indeed, the amplitude \eqref{eq:causal_amplitude} can be seen as the analog of the Hadamard propagator, which is a symmetrized 2-point correlation function solving the Klein-Gordon equation. However, to compute the transition probability between two positions and times, one must use the Feynman propagator, which is time-ordered Green-function of the Klein-Gordon equation. Both propagators can be computed using the path integral method. It is then shown that they differ by a different range of integration. In the Feynman propagator, one only sums over these trajectories evolve forward in time (time-ordered), which appears as a restriction of the range of the path-integral giving the Hadamard function.

To put it in a nutshell, the choice of including the causal terms or not depends on what we want to compute: a projector on the physical Hilbert space or an evolution operator.


\section{BF theory}
\label{sec:BF}

As a first step towards spin-foams, let's consider discrete BF theory. It is a topological theory, so we do not expect any causal structure to arise, but it makes use of an orientation structure which is worth looking at as a warm-up.

\subsection{Discrete BF theory}

Following \cite{Baez:1999sr}, the discretization of BF theory is done over a 2-complex $\mathcal{C}$. The variables are one group element $g_e \in G$ per edge $e\in \mathcal{C}$. Then the amplitude is defined as 
\begin{equation}
    \label{eq:Z_BF}
    Z_\mathcal{C} \overset{\text{def}}= \int [\dd g_e] \prod_f \delta \left( U_f (g_e) \right).
\end{equation}
There is one integral per edge,
$\dd g_e$ is the Haar measure over $G$, $\delta$ is the Dirac $\delta$-function over $G$ and the product is carried over all the faces $f$ of $\mathcal{C}$. Moreover we define the circular product
\begin{equation}
    U_f (g_e) \overset{\text{def}}= \prod_{e \in f}^\circlearrowleft g_e,
\end{equation}
where the product is made over the edges $e$ surrounding $f$. Although sometimes overlooked, we want to draw attention to the orientation structure which is required to define correctly the circular product. We need the following additional structure over $\mathcal{C}$:
\begin{enumerate}
    \item a distinguished edge to each face, that serves as a starting point in the product;
    \item an orientation to each face, that tells the order of the following edges.
\end{enumerate}
Although this structure is required to define $U_f$, $\delta(U_f)$ doesn't actually depend on it, due to the invariance of the $\delta$-function under inversion and cyclic permutation.

It is common to rewrite $Z_\mathcal{C}$ by splitting the $\delta$-function into a sum over the irreducible representations (irreps) of $G$:
\begin{equation}
    \label{eq:delta_expansion}
    \delta(U) = \sum_\rho \dim \rho \, \Tr \rho(U)
\end{equation}
Then, for each edge, the integral over the group element $g$ can be rewritten as a sum over intertwiners, which formally reads
\begin{equation}
    \label{eq:sum_iota}
    \int \dd g \bigotimes_{f \in e}^\circlearrowleft \rho_f(g)   = \sum_{\iota} \iota  \iota^*.
\end{equation}
Again, the definition of the circular tensor product requires us to introduce additional structure:
\begin{enumerate} \setcounter{enumi}{2}
    \item a distinguished face to each edge, that serves as a starting point in the tensor product;
    \item an orientation of the faces around each edge, which can be thought of as an arrow on the edge (with the right-hand convention to turn around for instance).
\end{enumerate}
Of course, $Z_\mathcal{C}$ remains blind to this structure. The amplitude finally becomes:
\begin{equation}
    \label{eq:amplitude_sums}
    Z_\mathcal{C} = \sum_\rho \sum_\iota \prod_f \dim \rho_f \prod_v A_v.
\end{equation}
The sum in $\rho$ (resp. $\iota$) is made over all the possible labeling of the faces (resp. edges) by irreps (resp. intertwiners). The \textit{vertex amplitude} $A_v$ is a function of the irreps $\rho_f$ and intertwiners $\iota_e$ attached to the faces and edges surrounding a vertex $v$.

The four orientation structures just introduced enter the computation of $A_v$. These structures lie on the edges and faces of the 2-complex. Motivated by our analysis in the previous sections, we want to investigate the idea that causality of spin-foams could arise from the breaking of the invariance of $Z_\mathcal{C}$ with respect to the orientation of $\mathcal{C}$.

\subsection{Ponzano-Regge model}

To proceed concretely, let's consider the simple example of the \textit{Ponzano-Regge model} \cite{Ponzano:1968a}, for which $\mathcal{C}$ is dual to a 3-dimensional simplicial complex $\Delta$, and $G = SU(2)$. In this case, the irreps are labeled by spins $j \in \mathbb{N}/2$ and there is no sum over the intertwiners (because it is unique). The vertex amplitude can be nicely represented pictorially as a graph where each node stands for an adjacent edge and each link for a face in-between. The \textit{vertex graph} takes typically the following form:
\begin{equation}
\label{eq:tetrahedron}
    \begin{overpic}[width = 0.35 \textwidth]{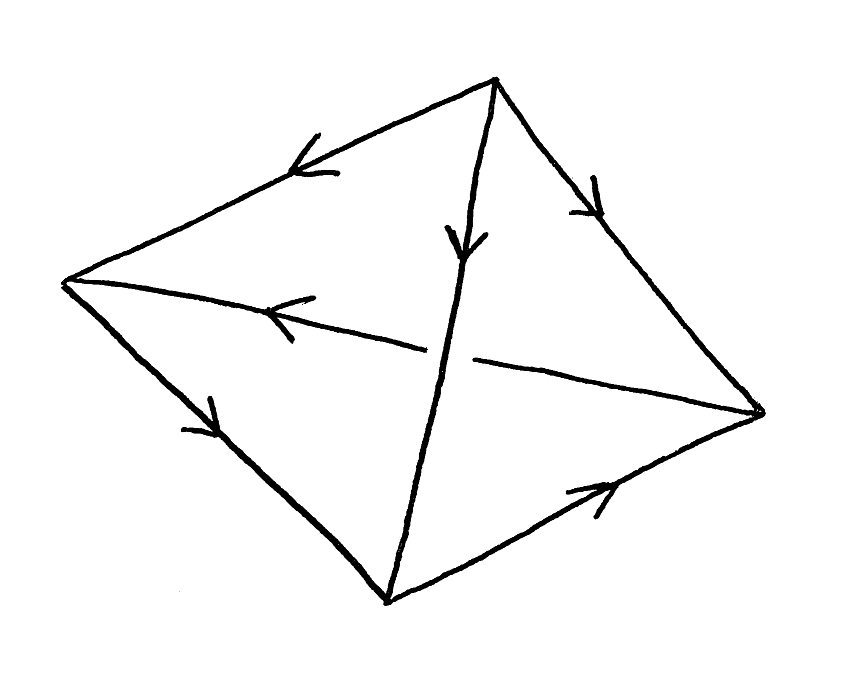}
    \put (0,48) {+}
    \put (45,8) {--}
    \put (58,73) {+}
    \put (90,30) {--}
    \put (30,70) {$j_1$}
    \put (48,57) {$j_2$}
    \put (72,55) {$j_3$}
    \put (36,35) {$j_5$}
    \put (20,20) {$j_6$}
    \put (80,20) {$j_4$}
    \end{overpic}
\end{equation}
The arrows on the links are induced by the orientation of the faces and the signs on the nodes are induced by the orientation of the edges (+ for incoming). Any combination of arrows and signs can be found, but the topology of the graph is the same for every vertex. The labels $j$ on the links are inherited from the irreps on the faces.\\
The graphical calculus is defined by the following rules:
\begin{enumerate}
    \item To each link $l$, associate a variable $m_l$ that will be summed over;
    \item The 3jm-Wigner symbol\footnote{We refer to \cite{Martin-Dussaud:2019ypf} for an introduction to the mathematical material used in this section.} is associated to the following nodes:
    \begin{equation}
        \begin{split}
        \begin{pmatrix}
        j_1 & j_2 & j_3 \\
        m_1 & m_2 & m_3
        \end{pmatrix}
        &=
        \begin{array}{c}
        \begin{overpic}[width = 0.15 \textwidth]{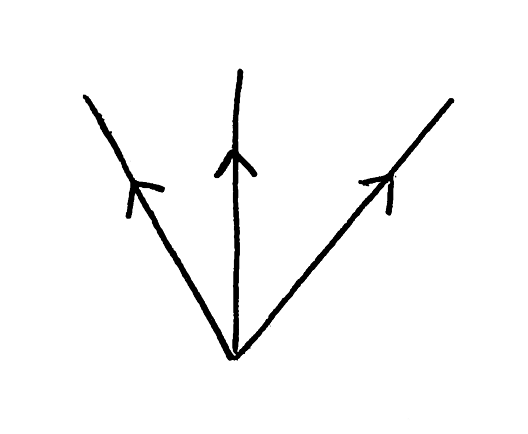}
        \put (40,0) {+}
        \put (10,48) {$j_3$}
        \put (50,60) {$j_2$}
        \put (80,40) {$j_1$}
        \end{overpic}
        \end{array} 
        =
        \begin{array}{c}
        \begin{overpic}[width = 0.15 \textwidth]{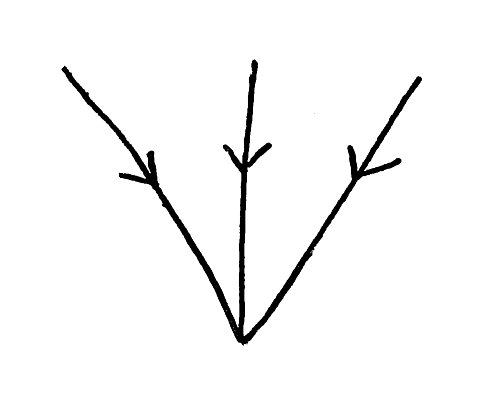}
        \put (40,0) {--}
        \put (10,48) {$j_1$}
        \put (55,60) {$j_2$}
        \put (80,40) {$j_3$}
        \end{overpic}
        \end{array}
        \end{split}
    \end{equation}
    The sign on the node indicates the sense in which the attached links shall be read.
    \item If an arrow is reversed, replace in the formula above $m_l$ by $-m_l$ and multiply by $(-1)^{j_l-m_l}$, like
    \begin{equation}
        \begin{array}{c}
        \begin{overpic}[width = 0.15 \textwidth]{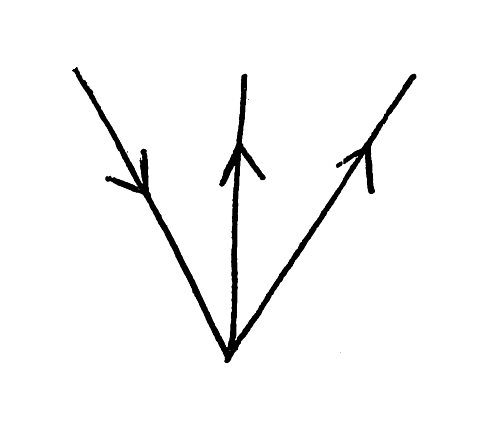}
        \put (40,0) {+}
        \put (10,48) {$j_3$}
        \put (50,60) {$j_2$}
        \put (80,40) {$j_1$}
        \end{overpic}
        \end{array}
        = (-1)^{j_3-m_3} \begin{pmatrix}
        j_1 & j_2 & j_3 \\
        m_1 & m_2 & -m_3
        \end{pmatrix}
    \end{equation}
    or
    \begin{equation}
        \begin{array}{c}
        \begin{overpic}[width = 0.15 \textwidth]{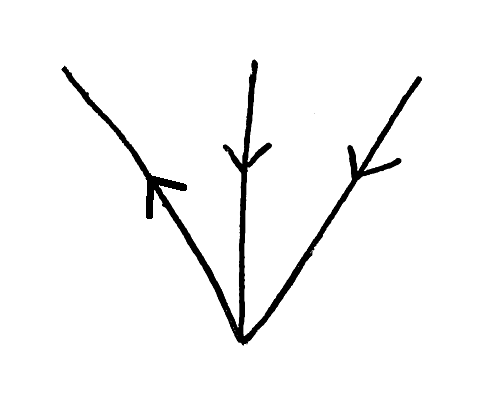}
        \put (40,0) {--}
        \put (10,48) {$j_1$}
        \put (55,60) {$j_2$}
        \put (80,40) {$j_3$}
        \end{overpic}
        \end{array}
        = (-1)^{j_1 - m_1} \begin{pmatrix}
        j_1 & j_2 & j_3 \\
        -m_1 & m_2 & m_3
        \end{pmatrix}
    \end{equation}
    A positive (resp. negative) node with an incoming (resp. outgoing) link corresponds to a counter-alignment of the face and the edge.
    \item Multiply all factors and sum over all $m_l$ from $-j_l$ to $j_l$ (integer steps).
\end{enumerate}
As an example, the graph \eqref{eq:tetrahedron} evaluates to 
\begin{multline}
    A_v = \sum_{m_i} (-1)^{j_4 - m_4 + j_1 - m_1} \begin{pmatrix}
        j_1 & j_2 & j_3 \\
        m_1 & m_2 & m_3
        \end{pmatrix} \\
        \times
        \begin{pmatrix}
        j_4 & j_5 & j_3 \\
        m_4 & -m_5 & m_3
        \end{pmatrix} 
        \begin{pmatrix}
        j_6 & j_2 & j_4 \\
        m_6 & m_2 & -m_4
        \end{pmatrix}
        \begin{pmatrix}
        j_6 & j_5 & j_1 \\
        m_6 & -m_5 & -m_1
        \end{pmatrix}
\end{multline}
The power of graphical calculus is apparent when comparing this cumbersome formula to the diagram \eqref{eq:tetrahedron}. Up to a sign, the vertex amplitude equals the 6j-symbol:
\begin{equation}
    A_v = \pm \begin{Bmatrix}
    j_1 & j_2 & j_3 \\
    j_4 & j_5 & j_6
    \end{Bmatrix}.
\end{equation}
The sign $\pm$ is a function of the spins $j_i$ and it depends on the orientation of the links and nodes. It matters when several vertices are glued together.

The importance of the Ponzano-Regge model was historically revealed by its semi-classical limit, which makes it a good candidate for 3D Euclidean quantum gravity \cite{Ponzano:1968a}. Indeed, the vertex amplitude admits a graphical representation as a tetrahedron depicted in \eqref{eq:tetrahedron}. This shape initially expresses the invariance of the 6j-symbol under the action of the tetrahedral group. But it turns out that it also carries a deeper geometric meaning when the labels $j_i$ are interpreted as the edge lengths of the tetrahedron. Denoting $V$ the volume of this tetrahedron, one can prove the following behavior for the vertex amplitude $A_v(\lambda j_i)$ when $\lambda \to \infty$:
\begin{equation}
\label{eq:classical_limit_PR}
    \begin{Bmatrix}
    \lambda j_1 & \lambda j_2 & \lambda j_3 \\
    \lambda j_4 & \lambda j_5 & \lambda j_6
    \end{Bmatrix} \sim \frac{1}{4\sqrt{3 \pi \lambda^3 V}} \left( e^{i S} + e^{-i S} \right)
\end{equation}
with the action
\begin{equation}
    S \overset{\text{def}}= \sum_i \left( \lambda j_i + \frac{1}{2} \right) \xi_i + \frac{\pi}{4} 
\end{equation}
with $\xi_i$ the exterior dihedral angle along the edge $i$ \cite{Roberts:1998zka}.

Graphical calculus also clarifies how the orientation enters the computation of the vertex amplitude. The invariance of the total amplitude $Z_\mathcal{C}$ under changes of orientation of the edges and faces is checked in appendix \ref{sec:orientation_invariance}. To be precise, the invariance is only true for faces and edges which lie in the bulk of $\mathcal{C}$. In general, $\mathcal{C}$ is bounded by some 3-valent graph $\Gamma$ over which an orientation is induced by $\mathcal{C}$. The total amplitude $Z_\mathcal{C}$ is sensitive to the orientation of $\Gamma$. The case is similar to the amplitude defined by equation \eqref{eq:causal_amplitude}. Although topological in the bulk, BF theory is non-trivial on the boundary. The boundary orientation provides a prototype of boundary causal structure.

A change of boundary orientation affects the value of $Z_\mathcal{C}$ in a simple way:
\begin{itemize}
    \item A flip of a link-orientation brings a global factor $(-1)^{2j}$ if the two-endpoints carry opposite signs, none otherwise.
    \item A flip of a node-orientation brings an overall factor $(-1)^{j_1+j_2+j_3}$.
\end{itemize}
As for Feynman diagrams, this simple way of modifying the causal structure of the boundary can be understood as a \textit{crossing symmetry}.

\subsection{Causal Ponzano-Regge model}

Now we want to go further and suggest a way to break the orientation-invariance in the bulk of the Ponzano-Regge model. A proposal of this kind can be found in \cite{Oriti:2006wq}, starting from a formulation of causality in terms of the flux variables of the discretized BF theory. Here, we adopt a different strategy, using the spin representation, motivated by the semi-classical limit. It exemplifies the strategy that will be adopted in the next section for the EPRL model \cite{Engle:2007wy}.

Consider $\mathcal{H}_j$, the spin-j irreducible representation of $SU(2)$, with the canonical basis $\ket{jm}$. The coherent states are defined as
\begin{equation}
    \ket{j,z} \overset{\text{def}}= u(z) \ket{j,-j}
\end{equation}
with $u : \mathbb{C}^2 \to SU(2)$ a (well-chosen) surjective map. Then, the intertwiner can be written as the state
\begin{equation}
    \ket{\iota} = \frac{1}{N} \int \dd g \, g \cdot \bigotimes_{i=1}^3 \ket{j_i, z_i} 
\end{equation}
with $N = N(j_i,z_i)$ such that $\braket{\iota}=1$. Up to a global phase, the vertex amplitude is
\begin{equation}\label{eq:A_v_K}
     A_v = \frac{1}{N} \int_{SU(2)} [\dd g_n] \prod_l  K_l(g_{t_l}, g_{s_l}) .
\end{equation}
$t_l$ and $s_l$ are respectively the source and the target of $l$. The wedge amplitude is defined as
\begin{equation}
     K (g_t, g_s) \overset{\text{def}}= \matrixel{j,z_t}{g^{-1}_t g_s }{j, z_s} = \matrixel{z_t}{g^{-1}_t g_s }{z_s}^{2j}.
\end{equation}
For the purpose of studying the semi-classical limit, it is convenient to write the wedge amplitude as
\begin{equation}
    K (g_t, g_s) = e^{2j (\log r  + i \theta)}
\end{equation}
with $r \in \mathbb{R}^+$ and $\theta \in (-\pi,\pi]$.

Now, to introduce a notion of causality, we can force some orientation structure to appear artificially: this is done here by writing the identity
\begin{equation}
     K (g_t, g_s) = \sum_{\varepsilon \in \{1,-1\}} \Theta(\varepsilon \theta ) \, e^{2j (\log r  + i \theta)},
\end{equation}
where $\Theta$ is the step function and we are summing over the signs $\varepsilon$. We then define the causal wedge amplitude as
\begin{equation}
     K^\varepsilon (g_t, g_s) \overset{\text{def}}= \Theta(\varepsilon \theta ) \, e^{2j (\log r  + i \theta)}.
\end{equation}
for some choice of $\varepsilon$, which can be understood as a choice of wedge orientation. Given one orientation $\varepsilon_l$ per wedge $l$, the causal vertex amplitude $A_v^{\varepsilon}$ is defined by replacing the wedge amplitude $K_l$ by its causal alternative $K_l^{\varepsilon_l}$ in equation \eqref{eq:A_v_K}. The BF vertex amplitude is recovered as 
\begin{equation}
    A_v = \sum_{[\varepsilon_l]} A_v^{\varepsilon_l},
\end{equation}
where the sum is made over all possible sign-assignation to the wedges. There is a total of $2^6$ such configurations. This sum introduces a partition of the range of integration of \eqref{eq:A_v_K} into as many sectors. In the semi-classical limit, only two sectors survive, as it appears in \eqref{eq:classical_limit_PR}. Since the exterior dihedral angles of any tetrahedron are always such that $\sin \xi \leq 0$, the two sectors are when $\varepsilon_l$ are either all positive or all negative. Starting with the causal vertex amplitude with all $\varepsilon_l$ negative thus leads to the asymptotic limit
\begin{equation}
    A^-_v \sim \frac{1}{4\sqrt{3 \pi V}}  e^{i S}.
\end{equation}
This provides a toy model for the appearance of causality, which we will apply to the EPRL model.

\subsection{$\{15j\}$ BF theory}

Before moving to the EPRL model, let us look at BF theory in 4 dimensions. The main difference with respect to the $3d$ case comes from the fact that $\Delta$ is a 4-dimensional simplicial complex and so the amplitude $Z_\mathcal{C}$ includes a sum over the intertwiners. The graphical representation of the intertwiners requires the introduction of an additional structure:
\begin{enumerate} \setcounter{enumi}{4}
    \item at each edge, the surrounding faces are partitioned into two sets of two (there exist three such partitions);
    \item these two sets are ordered (e.g. called left and right).
\end{enumerate}
Then the vertex amplitude is represented by a pentagram like
\begin{equation}
    \label{eq:pentagram}
    \begin{overpic}[width = 0.4 \textwidth]{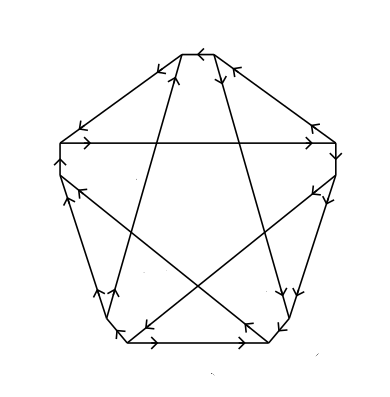}
    \put (67,10) {--}
    \put (73,17) {--}
    \put (25,76) {$j_1$}
    \put (15,35) {$j_2$}
    \put (50,8) {$j_3$}
    \put (80,35) {$j_4$}
    \put (70,80) {$j_5$}
    \put (49,59) {$j_6$}
    \put (38,50) {$j_7$}
    \put (42,36) {$j_8$}
    \put (55,39) {$j_9$}
    \put (57,50) {$j_{10}$}
    \put (48,90) {$\iota_1$}
        \put (43,90) {+}
        \put (53,90) {+}
    \put (8,59) {$\iota_2$}
        \put (10,64) {--}
        \put (10,54) {--}
    \put (25,12) {$\iota_3$}
        \put (22,17) {+}
        \put (30,9) {+}
    \put (73,12) {$\iota_4$}
    \put (87,59) {$\iota_5$}
        \put (87,54) {+}
        \put (87,64) {+}
    \end{overpic}
\end{equation}
$\iota \in \mathbb{N}/2$ is labeling the intertwiners. It is surrounded by two positive (resp. negative) nodes, when the edge is incoming (resp. outgoing). When the nodes are positive (resp. negative), the arrow goes from the right set to the left set (resp. the other way around). One can check with the rules above, that the overall amplitude $Z_\mathcal{C}$ is insensitive to the additional structure introduced. The same procedure as before can be used to define the causal vertex amplitude $A_v^\varepsilon$.

\medskip
To sum up, discrete BF theory is defined over a 2-complex $\mathcal{C}$ with a bunch of auxiliary orientation structures. The value of the partition function $Z_\mathcal{C}$ is sensitive to the boundary values of these structures, but not in the bulk. However, we have made a proposal to select a causal structure in the bulk as well.

\section{EPRL model and its causal structure}
\label{sec:EPRL}

General relativity can be formulated in a language close to the one of BF theory. The essential difference is that GR has local degrees of freedom, which mathematically arise from the implementation of constraints on the BF variables. Spin-foam models build upon this insight and consist in a weak implementation of these constraints in the discrete BF theory.

The appearance of the local degrees of freedom shows up in the breaking of the topological invariance of BF theory. The question then arises if these constraints induce also a causal structure on the 2-complex. The analysis of the previous sections suggests that such a causal structure can arise as an orientation structure on the edges or the wedges. Imposing constraints on these orientations then breaks the bulk orientation invariance of BF theory discussed above In this section, we propose such a construction for the EPRL model.

\subsection{Lorentzian EPRL model}

The EPRL model \cite{Engle:2007wy}, as formulated in \cite{Rovelli:2011eq}, is defined over a 2-complex $\mathcal{C}$ by the following partition function:
\begin{equation}
    Z_\mathcal{C} = \int_{SU(2)} [\dd h_w ] \, \prod_f \delta(U_f) \prod_v A_v(h_w).
\end{equation}
The integral is made over the variables $h_w \in SU(2)$ associated to each wedge $w$ in the bulk of $\mathcal{C}$. $Z_\mathcal{C}$ is a function of the variables $h_l \in SU(2)$ associated to each link of the boundary graph $\Gamma$. Similarly to BF theory, the precise definition requires additional structure on $\mathcal{C}$:
\begin{enumerate}
    \item a starting wedge per each face;
    \item an orientation per each face;
    \item an orientation to each wedge, i.e. each wedge $w$ has a source edge $s_w$ and a target edge $t_w$;
    \item a distinguished edge $E_v$ per each vertex $v$.
\end{enumerate}
Then, $U_f$ is defined as the circular product
\begin{equation}
    U_f (h_w) \overset{\text{def}}= \prod_{w \in f}^\circlearrowleft h_w
\end{equation}
that starts with the starting wedge of $f$, circulates in the sense given by the orientation of $f$, and each $h_w$ is inverted when the orientation of the wedge does not match with the orientation of the face. Besides, the vertex amplitude is
\begin{equation}
\label{eq:A_v_EPRL}
    A_v(h_w) = \int_{SL_2(\mathbb{C})} [\dd g_e ] \, \delta(g_{E_v}) \, \prod_{w \in v} K(h_w,g_{s_w} g_{t_w}^{-1})
\end{equation}
where there is one integration over $SL_2(\mathbb{C})$ per each edge surrounding $v$. The $\delta(g_{E_v})$ is only here to make the integral finite, but the value of $A_v$ is actually independent of the choice of $E_v$. The wedge amplitude $K$ is a function over $SU(2) \times SL_2(\mathbb{C})$ given by
\begin{equation}
    K(h,g) = \sum_j (2j+1)^2 \int_{SU(2)} \dd k \, \overline{\chi^j (hk)} \, \chi^{\gamma j,j}(kg),
     \label{eq:EPRL-K-def}
\end{equation}
with $\chi^j$ and $\chi^{p,k}$ respectively the characters of $SU(2)$ and $SL_2(\mathbb{C})$, and $\gamma \in \mathbb{R}$ the Barbero-Immirzi parameter.

The success of the EPRL model lies in its semi-classical limit, which was studied in \cite{Barrett:2009mw,Dona:2020yao} for a 2-complex $\mathcal{C}$ made of a single vertex dual to a Lorentzian 4-simplex $\sigma$. In this case, the partition function reduces to a single vertex amplitude $A_v(h_w)$, function of one $SU(2)$ element $h_w$ per each of the 10 links of the boundary graph. The boundary of $\sigma$ is made of 5 tetrahedra which can be described by the areas $j$ and the normals $\vec n$ to their faces. Then, the kinematics of loop quantum gravity prescribes a construction of a coherent state $\Psi_{j,\vec n}(h_w)$ which is ``peaked'' on the boundary geometry of $\sigma$. In the limit of large areas, $\lambda \to \infty$,
\begin{equation}
\label{eq:asymptotic_EPRL}
\begin{split}
    \braket{\Psi_{\lambda j, \vec n}}{A_v} &\overset{\text{def}}= \int_{SU(2)} [\dd h_w] \Psi^*_{j,\vec n}(h_w) A_v(h_w) \\
    &\sim \frac{1}{\lambda^{12}} \left( N_\sigma e^{i \lambda S_R} + N_{P\sigma} e^{- i \lambda S_R} \right)
\end{split}
\end{equation}
with $N_\sigma, N_{P \sigma} \in \mathbb{R}$ and $S_R$ is the Lorentzian Regge action of $\sigma$. This result takes  a form similar to \eqref{eq:classical_limit_PR} for the semiclassical limit of the Ponzano-Regge model. 

The EPRL model is blind to the bulk orientation structure that has been introduced to define it. Indeed, changing the starting wedge or the orientation of a face $f$ changes the value of $U_f$ but not of $\delta(U_f)$. Moreover, reversing the orientation of a wedge $w$ changes both $U_f$ and $A_v$. In $U_f$ it replaces $h_w$ by $h_w^{-1}$. In $A_v$, it interchanges $s_w$ and $t_w$, and so $K(h_w,g_{s_w} g_{t_w}^{-1})$ becomes $K(h_w,g_{t_w} g_{s_w}^{-1})$, which is easily shown to be equal to $K(h_w^{-1},g_{s_w} g_{t_w}^{-1})$. A change of variables $h_w \longrightarrow h_w^{-1}$ within the integral finally proves that the value of $Z_\mathcal{C}$ remains unchanged. 

The model is nevertheless sensitive to the orientation structure on the boundary. This structure consists only of the orientation of the links on the boundary over which the variables $h_l$ live. In the light of our preceding analysis, we expect this structure to carry causal information.

\subsection{Causal structures in the EPRL model}

Analogously to our previous toy model for BF theory, one can break the orientation invariance in the bulk as follows. First, one performs the integral over $k\in SU(2)$ in \eqref{eq:EPRL-K-def}, which yields
\begin{equation}
    K(h,g) = \sum_{j,m,n} (2j+1) \overline{D^j_{mn}(h) }   D^{\gamma j , j }_{jmjn}(g).
\end{equation}
One can switch from the magnetic basis $\ket{jm}$ to the overcomplete coherent states basis $\ket{j,z}$, by introducing a resolution of the identity
\begin{equation}
    \mathds{1}_j = \frac{2j+1}{\pi} \int_\Gamma \frac{\Omega(z)}{\|z\|^{4}} \dyad{j,z}
\end{equation}
with the measure
\begin{equation}
    \Omega(z) \overset{\text{def}}= \frac{i}{2} (z_0 \dd z_1 - z_1 \dd z_0) \wedge (z_0^* \dd z_1^* - z_1^* \dd z_0^*)\,.
\end{equation}
Here $\Gamma$ is the image of a path $\mathbb{C}P^1 \to \mathbb{C}^2$ that crosses each vector line of $\mathbb{C}^2$ once and only once\footnote{More abstractly, it can be regarded as a section of the Hopf bundle.}. 
One gets
\begin{equation}
    K (h,g) = \sum_j \frac{(2j+1)^3}{\pi^2} \int_{\Gamma'\times \Gamma''} \frac{\Omega( z') \Omega( z'')}{\|z'\|^4 \|z''\|^4} \matrixel{j,z'}{h^\dagger}{j,z''} \matrixel{\gamma j , j, j,z''}{D^{\gamma j, j}(g)}{\gamma j,j,j,z'}.
\end{equation}
Following \cite{Barrett:2009mw,Dona:2020yao}, the $SL_2(\mathbb{C})$ matrix element appearing above can be expressed in terms of an auxiliary $\mathbb{C}P^1$ as
\begin{equation}
    \matrixel{\gamma j,j,j,z''}{D^{\gamma j, j}(g)}{\gamma j,j,j,z'} = \frac{2j+1}{\pi} \int_\Gamma     \frac{\Omega(\zeta)}{\|\zeta\|^4}\; \mathscr{A}\; e^{i S_\gamma}\,,
\end{equation}
where
\begin{equation}
    \mathscr{A} =  \frac{\braket{\zeta}{z''^*}^{2j} \braket{z'^*}{g^T \zeta}^{2j} e^{2 i j (\arg z''_1 - \arg z'_1)} }{\|z'\|^{2j} \|z''\|^{2j}  \| \zeta \|^{2j-2}  \| g^T \zeta \|^{2j +2 }} 
\end{equation}
and the wedge action
\begin{equation}
\label{eq:wedge_action}
    S_\gamma = \gamma j \log   \frac{\| g^T \zeta \|^2}{ \| \zeta \|^2}\,.
\end{equation}
In the semi-classical, the wedge action becomes proportional to the wedge dihedral angle. This suggests to define the causal wedge amplitude as
\begin{equation}
K_\varepsilon (h,g) = \sum_j \frac{(2j+1)^4}{\pi^3} \int 
    \frac{\Omega(\zeta)\Omega( z') \Omega( z'')}{\|\zeta\|^4 \|z'\|^4 \|z''\|^4}    \matrixel{z'}{h^\dagger}{z''}^{2j}\;\; \Theta(\varepsilon S_\gamma ) \,\mathscr{A}\, e^{i S_\gamma}.
    \label{eq:causal-K}
\end{equation}
The causal vertex amplitude $A_v^\varepsilon$ is then defined by replacing $K$ with $K_\varepsilon$ in equation \eqref{eq:A_v_EPRL}, for some choice of $\varepsilon$ on each wedge. This defines a causal EPRL model.

The epithet `causal' can be further motivated by showing indeed that the extra variable $\varepsilon \in \{ 1,-1 \}$ encodes a causal structure on the wedges, as described previously. 

First, the usual vertex amplitude is recovered by summing over all possible sign-assignation to wedges:
\begin{equation}
\label{eq:sectors_EPRL}
    A_v = \sum_{[\varepsilon_w]} A_v^{\varepsilon}.
\end{equation}
There are $2^{10}$ terms in the sum. If one interprets $\varepsilon$ as wedge orientations, then a configuration only properly defines a causal structure if the cycle condition \eqref{eq:cycle} is fulfilled. So, one can properly call $A_v^\varepsilon$ a ``causal vertex amplitude'' when the configuration $[\varepsilon_w]$ satisfies \eqref{eq:cycle}, for either choice of signature $\eta \in \{-1,1\}$.

Secondly, assuming that the $[\varepsilon_w]$ indeed defines a proper causal structure, the word ``causal'' for $A_v^\varepsilon$ is only deserved if, in the asymptotic limit, $\varepsilon$ really captures the causal orientation of the boundary state. More precisely, given a Lorentzian 4-simplex, it determines a set of $\tilde \varepsilon$ for each wedge, such that $\tilde \varepsilon_w = \text{sign} \, \theta_w$, where $\theta_w$ is the dihedral angle of the wedge $w$. Then, its boundary coherent state $\Psi_{j,\vec n}$, when contracted with $A_v^{\tilde \varepsilon}$, should have the expected classical limit, i.e.
\begin{equation}
\label{eq:asymptotic_causal_EPRL}
    \braket{\Psi_{\lambda j, \vec n}}{A_v^{\tilde \varepsilon}} \sim \frac{1}{\lambda^{12}} N_\sigma e^{i \lambda S_R} 
\end{equation}
One can check that this is the case. Indeed, $\braket{\Psi_{\lambda j, \vec n}}{A_v}$ is an integral over the variables $g,z,z'$. The sum \eqref{eq:sectors_EPRL} introduces a partition of the range of integration in a number of sectors $[\varepsilon_w]$ characterized by $\varepsilon_w S_\gamma(x_w) > 0$, where $x_w$ stands for all the variables $g,z,z'$ on the wedge $w$. In the asymptotic limit, the two terms of \eqref{eq:asymptotic_EPRL} arise from two stationary points which are related by a parity transformation. The asymptotic analysis of \cite{Barrett:2009mw} shows that one of them, denoted $\sigma$, is such that $\tilde \varepsilon_w S_\gamma(x^\sigma_w) > 0$. Then the other, denoted $P\sigma$, satisfies $S_\gamma (x^{P\sigma}_w) = - S_\gamma(x^\sigma_w)$, so that $P\sigma$ and $\sigma$ are in opposite sectors. By construction, $\braket{\Psi_{\lambda j, \vec n}}{A_v^{\tilde \varepsilon}}$ selects only the sector of $\sigma$.

Our analysis suggests to interpret the two saddle points of the semi-classical limit as related by a switch of signature convention $\eta$. This interpretation differs from the one of \cite{Barrett:2009mw} where the two configurations were presented as arising from two parity-related 4-simplices. It also differs from the interpretation of \cite{Engle:2011un,Engle:2015mra} as two sectors of the Plebanski formulation intertwined with dynamical orientations. All these interpretations are related but it is not completely clear in which precise sense.

The asymptotic behavior \eqref{eq:asymptotic_causal_EPRL} is the only criterion that constrains the definition of the causal wedge amplitude. So, the dichotomy operated by $\Theta(\varepsilon S_\gamma )$ is to some extent arbitrary, and other functions could work as well. Another choice of $K_\varepsilon$ would define a different quantum theory with the same classical limit. Our choice appears to us as the simplest one. Its technical properties will be discussed in a second article.

To summarize, the EPRL amplitude can be understood as the sum of three contributions:
\begin{equation}
    A_v = \sum_{\substack{[\varepsilon_w] \\ \text{causal} \ \eta = 1}} A_v^{\varepsilon} + \sum_{\substack{[\varepsilon_w] \\ \text{causal} \ \eta = -1}} A_v^{\varepsilon} + \sum_{\substack{[\varepsilon_w] \\ \text{spurious}}} A_v^{\varepsilon}
\end{equation}
The two first terms correspond to configurations of $\varepsilon_w$ which satisfy the causal constraint \eqref{eq:cycle}, respectively for the choice of signature $\eta=1$ and $\eta=-1$. The last term contains all other configurations of $\varepsilon_w$. That both signatures enter in the amplitude is not surprising because both choices are allowed at the classical level. The two terms of the asymptotics in \eqref{eq:asymptotic_EPRL} can then be understood as arising from two different choices of signature. This again is to be expected because the Einstein-Hilbert action is such that
\begin{equation}
    S_{EH}[-g_{\mu \nu}] = - S_{EH}[g_{\mu \nu}],
\end{equation}
so that reverting the signature sends $S$ to $-S$.

Barrett et al. \cite{Barrett:2009gg} have also shown that the amplitude is not exponentially suppressed when the boundary is compatible with an Euclidean geometry in the bulk. In such a case, the group element $g$ at the critical point belongs to $SU(2)$ so that the wedge action \eqref{eq:wedge_action} vanishes. So the Euclidean configurations cannot be identified with a configuration of wedge orientations using our prescription.

When considering a full spin-foam, we obtain the following decomposition
\begin{equation}
    Z_\mathcal{C} = \sum_{\substack{[\varepsilon_w] \\ \text{causal} \ \eta=1}} Z^{\varepsilon_w}_\mathcal{C} + \sum_{\substack{[\varepsilon_w] \\ \text{causal} \ \eta=-1}} Z^{\varepsilon_w}_\mathcal{C} 
    + \sum_{\substack{[\eta_v] \\ \text{signature} \\ \text{changes}}}  \sum_{\substack{[\varepsilon_w] \\ \text{local} \\ \text{light-cones}}} Z^{\varepsilon_w}_\mathcal{C} + \sum_{\substack{[\varepsilon_w] \\ \text{spurious}}} Z^{\varepsilon_w}_\mathcal{C}\,.
\end{equation}
It then becomes an option to only consider some terms in the sum and thus partially or completely fix the causal structure in the bulk. The discussion of Sec.~\ref{sec:causal_path_integral} still holds. The choice to consider Feynman-like propagator or Hadamard-like function then depends on what is meant to be computed: a causal propagator or a projector on the physical Hilbert space. This conclusion answers some questions that were raised previously in the context of loop quantum cosmology \cite{Ashtekar:2010ve,Henderson:2010qd}.

The causal EPRL model is not a new model, but rather an interpretation of different components of the standard vertex amplitude in terms of causal structures. This interpretation is motivated \textit{a priori} by the understanding that discrete causal structures can be encoded on wedges, and \textit{a posteriori} by the asymptotics of the causal vertex. 

\section{Relation to earlier proposals}
\label{sec:relation_to_earlier_proposals}
We discuss how the proposal of the previous section relates to earlier results.   

\subsection{Livine-Oriti Barrett-Crane causal model}

Our approach is closely related to an earlier proposal on the implementation of causality by Livine and Oriti \cite{Livine:2002rh}. In the context of the Barrett-Crane model, the wedge amplitude is
\begin{equation}
    K^p(x_1,x_2) = \frac{2 \sin(\beta(x_1,x_2) \, p/2)}{p \sinh{\beta(x_1,x_2)}}\,,
\end{equation}
where $x_1$ and $x_2$ can be understood as the normals to two boundary tetrahedra and $\beta(x_1,x_2)$ is the Lorentzian angle in-between (see appendix \ref{sec:barrett_crane} for a complete definition of the symbols). The Livine-Oriti proposal consists in expanding the sine as
\begin{equation}
    K^p(x_1,x_2) = \frac{1}{p \sinh{\beta(x_1,x_2)}} \sum_{\varepsilon = \pm 1 } \varepsilon \, {e^{ i \varepsilon \beta(x_1,x_2) \, p/2}},
\end{equation}
interpreting  $\varepsilon$ as an orientation on the wedges and selecting one of the two sectors only. The implementation of causality discussed in this paper can be understood as a direct generalization of the Livine-Oriti proposal to the EPRL model. The non-trivial step introduced here is in the identification of how to introduce the splitting in \eqref{eq:causal-K}.

\subsection{Divergence and spikes}

The EPRL model and the Ponzano-Regge model both suffer from infrared divergences. In the latter case, the simplest example is provided by a triangulation $\Delta$ made of four tetrahedra subdividing a bigger tetrahedron. The dual 2-complex has 4 vertices and 10 faces. The four interior faces enclose a \textit{bubble}. The partition function is given by
\begin{equation}
    Z_\mathcal{C} = \sum_{j_i} \left(\prod_f (2j_f+1)\right) A_{v_1} A_{v_2} A_{v_3} A_{v_4}.
\end{equation}
The sum is made over the spins $j_i$ attached to the interior faces. The spins of the other faces are not summed over because they are fixed by the boundary conditions. The range of the sum for each interior face is a priori restricted by its neighboring faces according to the Clebsh-Gordan conditions. But the existence of a bubble implies that the sums are actually unbounded. This wouldn't be a problem if the vertex amplitude $A_v$ was decreasing fast enough for large spins, but such is not the case.

In the asymptotic limit, $A_v$ consists of two conjugate terms like
\begin{equation}
    A_v \sim A_v^+ + A_v^-,
\end{equation}
as shown in equation \eqref{eq:classical_limit_PR}.
The sign $\pm$ can be seen as an orientation of the tetrahedron dual to $v$. Thus,
\begin{equation}
    A_{v_1}A_{v_2}A_{v_3}A_{v_4} \sim \sum_{\varepsilon_i} A^{\varepsilon_1}_{v_1}A^{\varepsilon_2}_{v_2}A^{\varepsilon_3}_{v_3}A^{\varepsilon_4}_{v_4},
\end{equation}
where the sum is carried over all possible sign-assignation to the four vertices. In \cite{Christodoulou:2012af}, it is suggested that only some of the terms in this sum, dubbed ``spikes'', contribute to the divergence.  The spikes are specific configurations among those for which the light-cones are locally well-defined but with signature switches between the vertices. In other words, a wise selection of vertex orientation can cure the model from divergences. It is suggested that a similar behavior could cure the EPRL model as well.

Our proposed causal EPRL model \eqref{eq:causal-K} satisfies the requirements identified in \cite{Christodoulou:2012af}. However, there are two main differences with respect to their proposal: (i) We fix the orientation at the level of wedges, which is a finer scale than that of tetrahedra.  (ii) The orientation is fixed in the definition of the amplitude, while theirs only holds in the asymptotic limit.

\subsection{Engle's proper vertex}

The EPRL model has been criticized on the basis of its asymptotics, featuring the so-called \textit{cosine problem} \cite{Halliwell:1990qr}, \cite{Engle:2012yg,Vojinovic:2013faa}. The presence of two critical points is expected to cause problems when several vertices are considered \cite{Bianchi:2008ae}.
Engle has argued that the origin of this phenomenon is the fact that the EPRL model is built from discrete BF theory by imposing the simplicity constraint, but the latter is not strong enough as it admits three sectors out of the five of the Plebanski  formulation\cite{Engle:2011un,Engle:2015mra}. The proposal of a \textit{proper vertex} \cite{Engle:2015mra} includes  further constraints that restrict the model to the Einstein-Hilbert sector only. This is done by introducing in each wedge amplitude a spectral projector that concretely acts as a step-function $\Theta$. As a result, only one of the two terms in \eqref{eq:asymptotic_EPRL} is selected.

Our proposed causal vertex \eqref{eq:causal-K} shares a similar feature in that it amounts to the introduction of a step-function $\Theta$ on each wedge.\footnote{Note that, a priori, it is not clear that the restrictions introduced in the two proposals match away from the semi-classical limit. In fact, Engle's proper vertex introduces a step-function $\Theta$ on each wedge which depends on data on the full 4-simplex, while the step-function $\Theta$ in \eqref{eq:causal-K} is local on the wedge but includes the wedge orientations $\varepsilon_w$ as additional dynamical variables.} Yet, let us remark that the motivations are different. While Engle's proper vertex is motivated by the restriction to the Einstein-Hilbert sectors, and therefore to address the cosine problem, we are motivated by our analysis of the causal structure. It would be interesting to investigate thoroughly the relation between the sectors of Plebanski, the signature convention and the orientation of space-time, following the analysis of Immirzi in \cite{Immirzi:2016nnz}. Our analysis suggests that the appearence of the cosine in the asymptotics is a feature of the projector for the Hamiltonian constraint. Instead, when the spinfoam model is used as a causal propagator (as in Teitelboim's approach \cite{Teitelboim:1981ua,Livine:2002rh}), one selects only one class of causal configurations and finds only the contributions of the form $e^{i \lambda S_R}$ in the asymptotics.

\section{Conclusion}
The metric field of general relativity is almost fully determined by its causal structure. In this paper we have investigated the role played by the causal structure in spin-foam quantum gravity. We can summarise our main points as follows:\\
${}\quad-$ The notion of causality in general relativity encompasses two related but conceptually different notions: bare-causality and time-orientability.\\
${}\quad-$ There is a natural way to translate these notions to a simplicial complex.\\
${}\quad-$ The causal structure can be implemented on the dual 1-skeleton (edges). It can be seen as the combination of a causal set with a neighborhood relation.\\
${}\quad-$ It can also be encoded on the dual 2-skeleton (wedges) with a degeneracy of 2 that corresponds to a global time-reversal symmetry.\\
${}\quad-$ Starting from the set of all possible wedge orientations, the Lorentzian Regge action determines equations of motion whose solutions fix a proper causal structure.\\
${}\quad-$ The metric propagator can be written as a sum over all possible wedge orientations. By fixing the causal structure from the beginning, one defines a causal metric propagator, similar to the Feynman propagator.\\
${}\quad-$ The discrete BF theory naturally carries an orientation structure on the edges and faces, although it is blind to it in the bulk.\\
${}\quad-$ The discrete BF theory is sensitive to the orientation on the boundary. There are simple rules of crossing symmetry to go from one orientation to another.\\
${}\quad-$ There is a simple way to break the orientation invariance in the bulk, which provides a toy model to study causality in spin-foam models, \eqref{eq:causal-K} .\\
${}\quad-$ The EPRL amplitude can be regarded as a sum over all possible configurations of wedge orientations $\varepsilon_w$, which provide additional dynamical variables encoding the causal structure. Only a subset of it corresponds to properly causal configurations.\\
${}\quad-$ The causal EPRL vertex shares common traits with the Livine-Oriti causal version of the Barrett-Crane model and with Engle's proper vertex.\\
${}\quad-$ Whether one should use the causal or the full EPRL amplitude depends on what one wants to compute: a projector on the physical Hilbert space or a causal propagator.


\bigskip

\bigskip

\emph{Acknowledgments}. 
The authors thank Abhay Ashtekar, Pietro Don\`a, Francesco Gozzini, Alejandro Perez and Simone Speziale for insights and discussions during the course of this work. PMD thanks Alexandra Elbakyan for her help to access the scientific literature.

This work was made possible through the support of the ID\# 62312 grant from the John Templeton Foundation, as part of the project \href{https://www.templeton.org/grant/the-quantum-information-structure-of-spacetime-qiss-second-phase}{``The Quantum Information Structure of Spacetime'' (QISS)}. The opinions expressed in this work are those of the author(s) and do not necessarily reflect the views of the John Templeton Foundation. E.B.~acknowledges support from the National Science Foundation, Grant No.~PHY-2207851.

\appendix

\section{Orientation invariance of Ponzano-Regge model}
\label{sec:orientation_invariance}

The Ponzano-Regge model is defined over a 2-complex $\mathcal{C}$. To write the partition function like \eqref{eq:amplitude_sums}, it is necessary to introduce an orientation structure on the edges and faces. However, the value of $Z_\mathcal{C}$ is independent of this structure in the bulk. This can be checked as follows.

Consider an edge. Its orientation plays a role in the amplitudes of the two vertices that terminate the edge. Graphically, the contribution of the edge is
\begin{equation}
    \begin{overpic}[width = 0.2 \textwidth]{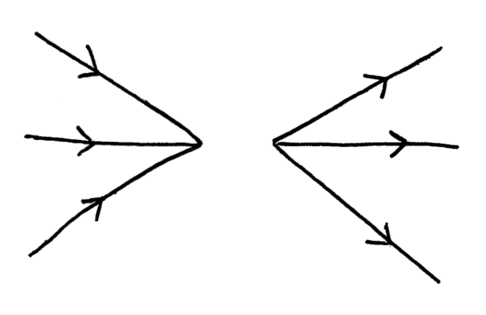}
    \put (48,35) {+}
    \put (43,30) {--}
    \put (-5,60) {$j_1$}
    \put (-5,35) {$j_2$}
    \put (-5,10) {$j_3$}
    \put (100,55) {$j_1$}
    \put (100,35) {$j_2$}
    \put (100,5) {$j_3$}
    \end{overpic}
\end{equation}
The face orientation, represented by the arrows, is actually irrelevant to the property that we are proving. What matters is that an edge spawns two nodes, one positive and one negative. Reversing the orientation amounts to switching the signs of both nodes. With the above rules, it is easy to check that 
\begin{equation}
\begin{array}{c}
\begin{overpic}[width = 0.2 \textwidth]{gfx/edge.png}
    \put (53,36) {--}
    \put (40,28) {+}
    \put (-5,60) {$j_1$}
    \put (-5,35) {$j_2$}
    \put (-5,10) {$j_3$}
    \put (100,55) {$j_1$}
    \put (100,35) {$j_2$}
    \put (100,5) {$j_3$}
    \end{overpic}
\end{array} \quad = \quad
    \begin{array}{c}
\begin{overpic}[width = 0.2 \textwidth]{gfx/edge.png}
    \put (48,36) {+}
    \put (43,30) {--}
    \put (-5,60) {$j_1$}
    \put (-5,35) {$j_2$}
    \put (-5,10) {$j_3$}
    \put (100,55) {$j_1$}
    \put (100,35) {$j_2$}
    \put (100,5) {$j_3$}
    \end{overpic}
\end{array}
\end{equation}
which proves the invariance under change of edge-orientation.

Now consider a face. Its orientation plays a role in the amplitudes of all the vertices around the face. For instance, in the case of a face labeled by $j$ and surrounded by three vertices, its orientation appears as an arrow on the $j$-link of three vertex amplitudes, like
\begin{equation}
    \begin{overpic}[width = 0.4 \textwidth]{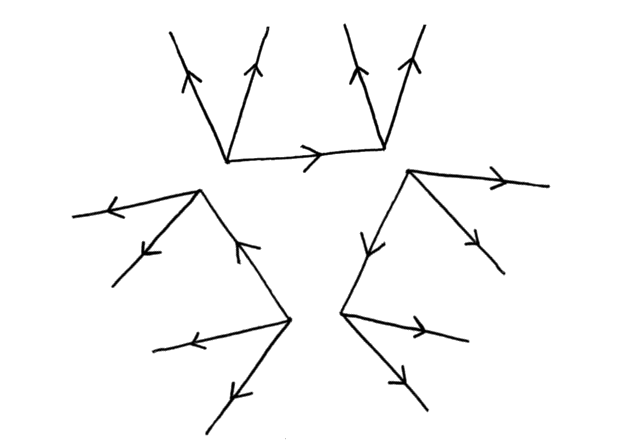}
    \put (33,38) {+}
    \put (30,44) {--}
    \put (60,45) {+}
    \put (60,41) {--}
    \put (44,16) {+}
    \put (49,21) {--}
    \put (50,50) {$j$}
    \put (32,25) {$j$}
    \put (60,25) {$j$}
    \put (46,33) {$f$}
    \end{overpic}
\end{equation}
Reversing the orientation of the face amounts to reversing the arrow on each of the links around. Now, using the above rules, it is easy to show that
\begin{enumerate}
    \item Reversing the arrow of a link in-between two positive or two negative nodes, amounts to multiplying by $(-1)^{2j}$.
    \item Reversing the arrow of a link in-between a positive and a negative link does not change the vertex amplitude.
\end{enumerate}
So overall, one gets a factor $(-1)^{2j}$
for each link surrounded by nodes of the same sign. To conclude, we need to prove the lemma that the number of such links around a face is always even. Indeed, choose a face and call $V$ the number of vertices around, and $D$ (resp. $S$)  the number of links around that have endpoint nodes with different (resp. the same) signs. We have $S+D=V$. Then notice that there are in total as many positive as negative nodes, so that the product of the signs of the nodes around a face is $(-1)^V$. The same quantity can be computed differently as $(-1)^{D}$. This shows that $V$ and $D$ have the same parity, which implies that $S$ is even and proves the lemma. As a conclusion, $Z_\mathcal{C}$ is indeed invariant under a change of face-orientation.

\section{Causal Barrett-Crane model}
\label{sec:barrett_crane}

In this appendix, we review the Livine-Oriti causal version of the Lorentzian Barrett-Crane spin-foam \cite{Livine:2002rh}.

\subsection{Lorentzian Barrett-Crane}

The Lorentzian Barrett-Crane model was introduced in \cite{Barrett:1999qw}. The 2-complex $\mathcal{C}$ is dual to a 4-dimensional simplicial complex $\Delta$ and each face $f$ is labeled by a positive number $p_f \in \mathbb{R}_+$. As formulated in \cite{Livine:2002rh}\footnote{Note that in  \cite{Livine:2002rh} the expressions are directly on the triangulation, instead of the dual picture.}, the partition function reads
\begin{equation}
    \label{eq:Z_Lorentzian_BC}
    Z_\mathcal{C} = \int_\mathbb{R^+} [\dd p_f] \, \prod_f p_f^2 \, \prod_e A_e \prod_v A_v. 
\end{equation}
The integral is carried over the labels $p_f$ for each face. The edge amplitude is given by
\begin{equation}
    A_e = \int_{(H^+)^2} \dd x_1 \dd x_2  \, \prod_{f \in e} K^{p_f}(x_1,x_2).
\end{equation}
The integration is carried over variables $x_1, x_2 \in H^+$ of the upper hyperboloid and the so-called kernel
\begin{equation}
    K^p(x_1,x_2) = \frac{2 \sin(\beta(x_1,x_2) \, p/2)}{p \sinh{\beta(x_1,x_2)}}
\end{equation}
with $\beta(x_1,x_2)$ the Lorentzian angle between $x_1$ and $x_2$:
\begin{equation}
    \beta(x_1,x_2) \overset{\text{def}}= \cosh^{-1}(x_1 \cdot x_2) \geq 0.
\end{equation}
Surrounding a vertex $v$, the edges are labeled with an index $i \in \{1,...,5\}$, and the faces are consistently labeled by a couple of such indices. Then the vertex amplitude is computed as
\begin{equation}
    A_v = \int_{(\mathcal{H}^+)^5} [dx_e] \, \delta(x_5 - 1) \, \prod_{(ij)} K^{p_{ij}}(x_i,x_j).
\end{equation}
It is striking to realize that the formulae are expressed without any reference to the orientation of $\mathcal{C}$. In this sense, the Lorentzian Barrett-Crane model is completely blind to the causal structure to the extent that it does not even take care of the causal structure on the boundary.

\subsection{Livine-Oriti causal model}

The latter fact was recognised as unsatisfactory by Livine and Oriti in \cite{Livine:2002rh}. Rewriting the kernel as
\begin{equation}
    K^p(x_1,x_2) = \frac{1}{p \sinh{\beta(x_1,x_2)}} \sum_{\varepsilon = \pm 1 } \varepsilon \, {e^{i \varepsilon \beta(x_1,x_2) \, p/2}}
\end{equation}
we see that the Lorentzian Barrett-Crane gives the amplitude:
\begin{equation}
    Z(\Delta) = \sum_{\varepsilon_f} \int_{(\mathcal{H}^+)^5} [dx_e] \, \delta(x_5 - 1) 
    \left( \prod_{f \in v}  \frac{\varepsilon_f}{p_f \sinh{\beta_f}}  \right) 
     e^{i \sum_f \varepsilon_f \beta_f \, p_f/2}.
     \label{eq:Z_1_vertex}
\end{equation}
where the sum is carried over all possible sign assignation to the faces $f$ around the vertex $v$. The $p_f$ are related to the area of the boundary triangle by
\begin{equation}
    p_f = \lambda A_f.
\end{equation}
In the semi-classical limit, when $\lambda \longrightarrow \infty$, the sum over $\varepsilon_f$ and the integral over $x_e$ in \eqref{eq:Z_1_vertex} are dominated by only two terms for which the variables describe the two only possible 4-simplices which have $p_f$ as areas  up to degenerate contributions.

The coexistence of two terms in the asymptotic limit can be interpreted as a time-reversal symmetry of the model. So Livine and Oriti proposed to change the vertex amplitude of the Barrett-Crane model by truncating the sum over $\varepsilon_f$ to keep only the term that matches the causal structure of the dual 4-simplicial complex. This defines a causal amplitude with only one term in the asymptotic limit. When $\Delta$ consists of many 4-simplices, the same truncation leads to the selection of a single causal structure on $\Delta$.


\end{document}